\documentclass[ejs]{imsart}

\RequirePackage[OT1]{fontenc}
\RequirePackage{amsthm,amsmath}
\RequirePackage{natbib}

\RequirePackage[colorlinks,citecolor=blue,urlcolor=blue]{hyperref}

\startlocaldefs
\numberwithin{equation}{section}
\theoremstyle{plain}

\endlocaldefs

\RequirePackage[OT1]{fontenc}
\RequirePackage{amsthm,amsmath}
\usepackage{enumerate} 
\usepackage{graphicx}
\usepackage{amsfonts}
\usepackage{bm,bbm} 
\usepackage[]{algorithm2e}
\usepackage{epstopdf}
\usepackage{caption}
\usepackage{subcaption}
\RequirePackage{color}
\usepackage{soul} 

\usepackage{multirow}

\usepackage{color}

\newcommand{\new}[1]{{#1}}
\newcommand{\remove}[1]{}

\RequirePackage{bm}

\usepackage[font = {it}]{caption}

\newcommand{\Proba}{\rm{Pr}}

\newcommand{\indic}{\mathbbm{I}}
\newcommand{\Msun}{M_{\odot}}

\newcommand{\btheta}{\boldsymbol{\theta}}

\newcommand{\msim}{m_{\text{sim}}}
\newcommand{\mobs}{m_{\text{obs}}}

\newcommand{\nsim}{n_{\text{sim}}}
\newcommand{\nobs}{n_{\text{obs}}}

\newcommand{\Cmin}{C_{\text{min}}}
\newcommand{\Cmax}{C_{\text{max}}}

\newcommand{\Mtot}{M_{\text{tot}}}

\newcommand{\Mmax}{M_{\text{max}}}
\newcommand{\Mmin}{M_{\text{min}}}

\begin{document}

\begin{frontmatter}
\title{A Preferential Attachment Model for the Stellar Initial Mass Function}
\runtitle{PA Model for IMF}

\begin{aug}
\author{\fnms{Jessi} \snm{Cisewski-Kehe}
\ead[label=e1]{jessica.cisewski@yale.edu}}
\address{Department of Statistics \& Data Science \\
Yale University \\
New Haven, CT 06511\\
\printead{e1}}

\author{\fnms{Grant} \snm{Weller}\ead[label=e2]{gweller57@gmail.com}}
\address{UnitedHealth Group Research \& Development \\
Minneapolis, MN 55430 \\
\printead{e2}}

\author{\fnms{Chad} \snm{Schafer}
\ead[label=e3]{cschafer@cmu.edu}
}
\address{Department of Statistics \& Data Science\\
Carnegie Mellon University \\
Pittsburgh, PA, 15213 \\
\printead{e3}\\
}

\runauthor{J. Cisewski-Kehe et al.}


\end{aug}

\begin{abstract}
Accurate specification of a likelihood function is becoming increasingly difficult in many inference problems 
in astronomy.
As sample sizes resulting from astronomical surveys continue to grow, 
deficiencies in the likelihood function lead to larger biases in key parameter estimates.
These deficiencies result from the oversimplification of the physical processes that
generated the data, and from the failure to account for observational limitations.
Unfortunately, realistic models often do not yield an analytical form for the likelihood.
The estimation of a stellar initial mass function (IMF) is an important example. The stellar IMF is the mass 
distribution of stars initially formed 
in a given cluster of stars, a population 
which is not directly observable due to stellar evolution and other disruptions and observational limitations of the
cluster. There are several difficulties with specifying a likelihood in this setting
since the physical processes and observational challenges result in measurable 
masses that cannot legitimately be considered independent draws from an IMF.
This work improves inference of the IMF by using an approximate Bayesian computation approach 
that both accounts for observational and astrophysical effects and incorporates
a physically-motivated model for star cluster formation.
The methodology is illustrated via a simulation study, demonstrating that the proposed approach can recover the true posterior in
realistic situations, and applied to observations from astrophysical simulation data.  \\
The published version of this manuscript is available at \url{https://doi.org/10.1214/19-EJS1556}.
\end{abstract}


\tableofcontents
\end{frontmatter}

\section{Introduction}
\label{introSec}
The Milky Way is home to billions of stars \citep{McMillan:2016uq}, many of which 
are members of \emph{stellar clusters} - gravitationally bound collections of stars. 
Stellar clusters are formed from low temperature and high density clouds of gas and dust 
called \emph{molecular clouds}, though there 
is uncertainty as to how the stars in a cluster form \citep{Beccari2017}. 
Each theory of star formation yields a different prediction for the distribution
of the masses of stars that initially formed in a cluster. Hence, it is of fundamental 
interest to estimate this distribution, referred to as the \emph{stellar initial mass function (IMF)},
and assess the validity of these competing theories.
\new{The IMF can be thought of as a continuous density describing the distribution of star masses that initially form in a stellar cluster.}
In fact, research advances in many areas of stellar, galactic, and extragalactic astronomy are at 
least somewhat reliant upon accurate understanding of the IMF \citep{bastian2010}.
For example, the IMF is a key
component of galaxy and stellar evolution and planet formation \citep{bally2005, bastian2010, Shetty2014}, 
along with chemical enrichment and abundance of core-collapse supernovae \citep{weisz13}.

There is also ongoing discussion surrounding the {\it universality} of the IMF, i.e.,
if a single IMF describes the generative distribution of stellar masses for all
star clusters \citep{bastian2010}. The consensus of the astronomical community is that the IMF is not universal, however, most of the observations had been consistent with universality \citep{kroupa2001, bastian2010, Ashworth2017}.
With further research and growing sample sizes, however, there is increased theoretical 
(e.g., \citealt{bonnell2006, Dib2010})
and observational
(e.g., \citealt{Treu2010, Dokkum:2010fk, Spiniello2014,Geha2013, Dib2017})
support for an IMF that can vary cluster to cluster. 

\cite{salpeter55} studied the evolutionary properties of certain 
populations of stars, and in the process defined the first IMF 
(which he called the ``original mass function'').  
This work put forth the now-classic model for the IMF, a power law 
with a finite upper bound equal to the physical maximum mass of a star that 
could form in a cluster \citep{salpeter55}.  
More recent studies continue to use this power law form for the IMF, especially for stars of mass greater than half that
of our sun (e.g., \citealt{Massey2003, bastian2010, DaRioEtAl2012, Lim2013, weisz13, Weisz:2015kx, Jose2017}).
Similar models have been proposed and used in the astronomical literature for inference of the 
stellar IMF; these will be discussed in the next section. 
The estimation of the parameters of these proposed models typically relies on the assumption that 
the observed stars in a stellar cluster form independently; more specifically, the assumption that the masses of the individual stars form independently. 
The proposed model in this work loosens the assumption of independence in order to explore one of several possible physical formation 
mechanisms of cluster formation\new{ and avoids specification of a parametric model form by using on a new simulation model}.

Despite this seemingly simple form of the power law model, the statistical challenges of estimating the 
IMF using \emph{observed} stars from a cluster are significant. Many of the limitations are related 
both to 
observational issues and to the adequate modeling of the evolution of a star cluster after the 
initial formation. For example, since stars of greater mass die more rapidly, the upper tail of the IMF is 
not observed in a cluster of sufficient age.  
Also, the death of massive stars can trigger additional star formation,
contaminating the lower end of the IMF with new stars \citep{Woosley2015}.
There are also issues related to missing lower-mass stars due to the sensitivity of the instruments.  
The observational astronomers will often estimate a \emph{completeness function} of an observed 
cluster, which is the probability of observing a star of a particular mass.  The completeness 
function is discussed in more detail below.

The observational limitations and the challenge of modeling cluster evolution make approximate Bayesian 
computation (ABC) appealing for estimation of the IMF, as ABC allows for relatively easy incorporation of
such effects. The difficulty of addressing these limitations is 
implied by the fact that observational effects are often ignored or accounted for in 
an ad hoc or unspecified manner (e.g., \citealt{DaRioEtAl2012, Ashworth2017, Jose2017, Kalari2018}), 
though \cite{weisz13} discuss how some observational limitations can be incorporated into their 
proposed Bayesian model.
A primary appeal of ABC for this application is the ability to incorporate more complex models for cluster
formation. 
Standard IMF models do not specify the process by which
a large mass of gas (the molecular cloud) transforms into a gravitationally bound collection of stars. 
ABC is based on a simple rejection-sampling approach, in which draws of model parameters from a prior 
distribution are fed through a simulation model to generate a sample of data. 
If the generated sample is ``close'' (based on an appropriately chosen metric) to the 
observed data, the prior draw that produced that generated sample is retained. The collection of
accepted parameter values comprise draws from an approximation to the posterior.
The simulations
(the \emph{forward model}) can include any of the complex processes that make it challenging to derive a
likelihood function for the observable data.

This situation is typical of inference challenges that arise in astronomy. See
\cite{schafer2012, AkeretEtAl2015, IshidaEtAl2015} for reviews.
Recent years have seen a
rapid increase in the use of ABC methods for estimation in this field, including
specific application to 
Milky way properties \citep{RobinEtAl2014},
strong lensing of galaxies \citep{Killedar2018,Birrer2017},
large scale structure of the Universe \citep{Hahn2017b},
estimating the redshift distribution \citep{Herbel2017},
galaxy evolution \citep{Hahn2017a},
weak lensing \citep{Peel2017,Lin2015},
exoplanets \citep{Parker2015},
galaxy morphology \citep{CameronPettitt2012},
and
supernovae \citep{WeyantEtAl2013}.

%


Hence, our motivation for using the proposed stochastic model for the stellar IMF includes both scientific and practical considerations. We model the observable data in a star cluster using a formation mechanism that incorporates realistic dependency in the masses of the stars. Further, this model generalizes commonly used IMF models, i.e., it can capture, but also distinguish, popular competing IMF model shapes. Flexible models of this type have great potential to test widely-held assumptions of more restrictive parametric forms, and eliminate the need for (often arbitrary) model selection exercises. Finally, the generative approach allows for the incorporation of observation effects and uncertainties within an ABC framework.

This paper is organized as follows. In Section \ref{sec:background}, background on the IMF along with inference challenges are presented along with an introduction of ABC. 
%
The proposed stochastic model for stellar formation is discussed in Section  \ref{PAmodelSection}. A simulation study, including an application of the proposed methodology to the estimation of the IMF of a realistic astrophysical simulation \citep{Bate2012}, can be found in Section \ref{sec:sim}. Finally,  \ref{discussionSec} provides a discussion.

\section{Background}
\label{sec:background}
\subsection{Stellar Initial Mass Function}

As noted above, \cite{salpeter55} introduced the power law model for the shape of the IMF for masses larger
than 0.5$\Msun$, where $\Msun$ is the mass of the Sun.
\cite{kroupa2001} extended the range of the IMF by proposing a three-part broken power law model 
over the range $0.01 \Msun < m < \Mmax$, where $\Mmax$ is the mass of the largest star that could form with nonzero probability.
This model postulates different forms for the IMF for stars of masses $0.01 \Msun < m < 0.08 \Msun$,
$0.08 \Msun < m < 0.5$ and $m > 0.5\Msun$.
\remove{Focusing on the upper part} \new{To illustrate the form of the IMF model of \cite{salpeter55}, consider the upper tail where $m > 0.5\Msun$}, and \remove{defining} \new{define} $\theta = (\alpha, \Mmax),$ \new{then} the probability density function for \remove{mass $x$} \new{ mass $m$} in the upper tail of the stellar IMF is assumed to be given by
\begin{align}
f_M(m \mid \theta) = cm^{-\alpha}\text{,}\;\; m \in [0.5\Msun, \Mmax]\text{,}
	\label{eq:imf}
\end{align}

\begin{figure}[htbp]
   \centering
\includegraphics[width = .5\textwidth]{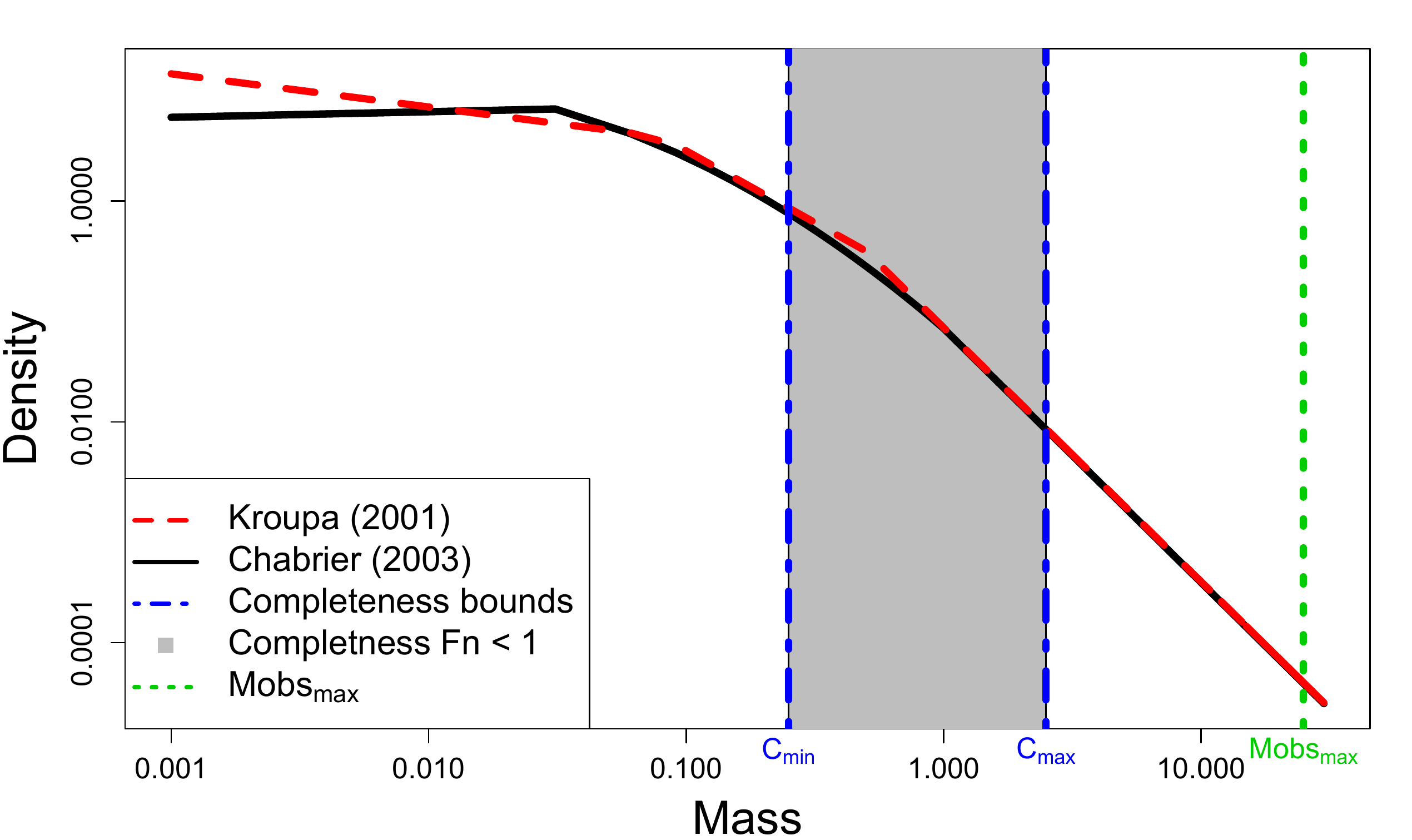} 
   \caption{The broken power law model of \cite{kroupa2001} (red, dashed) and the log-normal with a power law tail model of \cite{Chabrier:2003oq, Chabrier:2003om} (black, solid) are displayed along with vertical lines representing several observational challenges.  The blue vertical dotted-dashed lines indicate the range of values ($C_{\min}, C_{\max}$) on which the completeness function may be defined, and the vertical green dotted line indicates the maximum observable mass (Mobs$_{\max}$) due to the aging cutoff.  The observational challenges are discussed further in \S\ref{sec:obs_challenges}.}
   \label{fig:imf_models}
\end{figure}

\remove{$f_M(m \mid \theta) = cm^{-\alpha}\text{,}\;\; m \in [\Mmin, \Mmax]$}

\noindent where the constant $c$ is chosen such that $f_M$ is a valid probability density.  
\new{(For the form of the IMF model of \cite{kroupa2001}, see Equation~\eqref{eq:kroupa}.)}
Alternative models have been proposed that include log-normal distributions, joint power law and log-normal parts, and truncated exponential distributions  \citep{Chabrier:2003oq, Chabrier:2003om, chabrier2005,IMF50,bastian2010, OffnerEtAl2014}.  The \cite{kroupa2001} and \cite{Chabrier:2003oq, Chabrier:2003om} models are displayed in Figure~\ref{fig:imf_models} along with observational challenges discussed \S\ref{sec:observational}.
Power law distributions and log-normal distributions are closely related and may be the result of subtle differences in the underlying formation mechanism \citep{Mitzenmacher2004}.  
The IMF model we propose will include, as a special case, a family of formation mechanisms that generate
power law tails, but also allow for a wider range of tail behaviors (see \remove{Section} \new{Appendix}~\ref{sec:powerlaw}).

\subsubsection{Observational Challenges} \label{sec:observational}
Observing all stars comprising an IMF 
is not feasible, as the most massive stars ($m > 10 \Msun$) have lifetimes of 
only a few million years.
The lifetime of a star (the time it takes for the star to burn through its
hydrogen) depends strongly on its mass:~the most massive
stars have shorter lives due to the hotter temperatures they must maintain
to avoid collapse from the strong gravitational forces. In particular,
stellar life is 
approximately proportional to $m^{-\rho}$ 
where $\rho \approx 3$ (\citealt[p. 30]{hansen2004}, \citealt[p. 439]{Chaisson:2011}). 
Hence, the mass of the largest star observed in a given 
cluster depends on the cluster age.  

Furthermore, the IMF is estimated using a noisy, incomplete view of that cluster.  Whether or not a star is observed is dependent on several factors including its mass, its location in the cluster, and its neighbors.  Some of these factors are described by a data set's \emph{completeness function}, which quantifies a given star's probability of being observed.  This depends on its luminosity (i.e. intrinsic brightness) since it needs to be sufficiently bright to be observable; in particular, completeness depends on stellar flux in comparison with the flux limits of the observations.
There are also issues with {\it mass segregation}:~stars with lower mass tend to be on the edge of the cluster, while the most massive stars are often
found in the center \citep{weisz13}. Due to {\it stellar crowding} in the center, stars in this region can be more difficult to observe.  Additionally, binary stars (star systems consisting of a pair of stars) 
are difficult to distinguish from a single star, creating the potential for overstating the mass of an object and understating the number of stars in the cluster.

There are additional uncertainties involved in translating the actual observables  (e.g. photometric magnitudes) into a mass measurement; that is, the mass values for observable stars are only estimates. The {\it Hertzsprung-Russell (H-R) Diagram} is a classic visual summary of the distribution of the luminosity and temperature of a collection of stars. A typical H-R Diagram includes a {\it main sequence} of stars that trace a line from bright and hot stars to dim and cool stars. Stellar mass also evolves along this one-dimensional feature, and since luminosity and temperature are estimable, mass can thus also be estimated. The mass of binary stars can be determined via Kepler's Laws, and hence a {\it mass-luminosity relationship} can be fit to binaries and then extended to other stars on the main sequence. Unfortunately, luminosity and temperature are nontrivial to estimate, as corrections for effects such as {\it accretion} and {\it extinction} are required, along with an accurate estimate of the distance to the stars \citep{Da-Rio:2010aa}. 
The process is further complicated by the dependence of how these transformations are made on the {\it spectral type} of the star. Careful budgeting of the errors that accumulate is required in order to produce a reasonable error bars on mass estimates; \cite{Da-Rio:2010aa} utilize a Monte Carlo approach in which the errors in magnitudes are propagated forward through to uncertainties in the spectral type, the accretion and reddening corrections, and finally to an uncertainty on the mass.

\subsection{Approximate Bayesian Computation}

Standard approaches to Bayesian inference, either analytical or built on
MCMC, require the specification of a
likelihood function, $f(m \mid \btheta)$, with data $\mobs \in \mathcal D$, 
and parameter(s) $\btheta \in \Theta$. In many modern scientific inference 
problems, such as for some emerging models for the stellar IMF, the likelihood is too complicated to be derived or otherwise
specified. 
As noted previously, ABC provides an approximation to the posterior
without specifying a likelihood function, and instead relies on forward simulation of the data generating
process.

The basic algorithm for sampling from the ABC posterior is attributed to 
\cite{TavareEtAl1997} and \cite{PitchardEtAl1999}, used for applications to population 
genetics. The algorithm has three main steps which are repeated until a sufficiently
large sample is generated:
\emph{Step 1}, Sample $\btheta^*$ from the prior; \emph{Step 2}, 
Generate $\msim$ from the forward process assuming $\btheta^*$; 
 \emph{Step 3},  
Accept $\btheta^*$ if $\rho(\mobs, \msim) \leq \epsilon$, where $\rho(\cdot, \cdot)$ is a user-specified distance function and $\epsilon$ is a tuning parameter that should be close to 0.
This last step typically consists of comparing low-dimensional summary statistics generated
for the observed and simulated datasets. 
Adequate statistical and computational performance of ABC algorithms depends greatly on the
selection of such summary statistics
\citep{JoyceMarjoram2008,BlumFrancois2010, Blum2010, FearnheadPrangle2012, BlumEtAl2013}.

The basic ABC algorithm can be inefficient in cases where the parameter space is of moderate
or high dimension.
Hence, important adaptations of the basic ABC algorithm incorporate ideas of sequential sampling in 
order to improve the sampling efficiency \citep{MarjoramEtAl2003,SissonEtAl2007,beaumont2009, DelMoralEtAl2011}.  A nice overview of ABC can be found in \cite{MarinEtAl2012}.  
Here, we use a sequence of decreasing tolerances $\epsilon_{1:T} = (\epsilon_1, \ldots, \epsilon_T)$ 
with the tolerance $\epsilon_t$ shrinking until further reductions do not significantly affect the 
resulting ABC posterior.
The improvement in efficiency is due to the modification that happens after the first time step: 
instead of sampling from the prior distribution, the proposed $\btheta$ are drawn from the previous time step's ABC posterior.  Using this adaptive proposal distribution can help to improve the sampling efficiency.  The resulting draws, however, are not targeting the correct posterior, and so importance weights, $W_t$, are used to correct this discrepancy.


\section{Forward Model for the IMF}
\label{PAmodelSection}
Due to their simple interpretations, mathematical ease, and demonstrated consistency with observations, power law IMFs (or similar variants) have been widely adopted in the astronomy literature \citep{kroupa2012}; however, open questions remain about stellar formation processes.  The proposed forward model is a way to link a possible stellar formation process with the realized mass function (MF).  

\new{
One known underlying mechanism for producing data with power-law tails is based on {\em preferential attachment} (PA) \citep{Mitzenmacher2004}.  
The earliest PA model, the Yule-Simon process, was popularized by \cite{simon55}, and was originally used to model biological genera and word frequencies. Other PA models include the classic {\it Chinese restaurant process}
and its generalizations \citep{Bloem2017}.
Interest in PA models grew within the study network evolution
\citep{BarabasiAlbert1999}.
Such evolution is described by the {\it attachment function}, which
describes the probability that a node acquires an additional edge,
usually as an increasing function of its current degree.
Most of the work done on estimation of the attachment function
makes the assumption that observations are available regarding
the full or partial evolution of the network.
This includes the nonparametric methods of 
\cite{Jeong2003}, \cite{Newman2001}, and \cite{Pham2015};
the maximum likelihood approaches of 
\cite{Gomez2011}, \cite{Wan2017a}, \cite{Onodera2014}; and
the Bayesian approach (using MCMC) taken by \cite{Sheridan2012}.
\cite{Wan2017a} also describes 
an approximation to the MLE that can be utilized when only a snapshot view of
the network is available.
\cite{Wan2017b} uses a semiparametric approach to fit to the upper tail of the
network degree distribution. The focus is on how the estimator performs
under deviations from the linear PA model and the ``superstar" linear PA
model, in which one node to which most of the other nodes attach. Estimation 
is based on extreme value theory.
\cite{Kunegis2013} use a simple least squares method to estimate the exponent
in a nonlinear, but parametric, PA model.}

\new{
In what follows, the proposed data generating process will use ideas of PA
to model the the evolution of a star cluster. The ABC approach will be
well-suited to perform inference with this model, given its complexity and
the available data.
}

\subsection{Preferential attachment for the IMF} \label{sec:pa}
The formation of a star cluster is a complicated and turbulent process with different theories on the physical processes that lead to the origin of the stellar IMF \citep{chabrier2005, Bate2012, OffnerEtAl2014, Pokhrel:2018nr}.  It is generally understood that the molecular cloud fragments and then forms stellar cores with a distribution referred to as the \emph{core mass function} (CMF).  Whether evolution from the CMF to the IMF is random, deterministic, or something in between is debated \citep{OffnerEtAl2014}.  In the proposed work, we consider the case where star cores can increase in mass by accreting material from the surrounding cloud and a particular star's final mass can be affected by its neighbors through turbulence or dynamical interactions.
That is, rather than assuming that stellar masses in a cluster arise independently of each other, our PA model proposes a resource-limited mass accretion process between stellar cores whose ability to accumulate additional mass is a function of their existing masses. 
This dependence feature is particularly important for statistical inference, as models that assume independent observations of stellar masses are vulnerable to incorrect and misleading inference. 
Additionally, the mass of the largest star to form in a cluster is limited by the total cluster mass.

Our proposed stochastic model for stellar formation is as follows:~we first fix a total available cluster mass $\Mtot$. 
This quantity can be physically interpreted as the total mass available for stellar formation in a molecular cloud. 
At each time step $t = 1, 2, \ldots$, a random quantity of mass $m_t \sim \text{Exponential}(\lambda)$ enters the collection of stars; $M_{1,1} = m_1$ becomes the mass of the first star.
Subsequent masses entering the system form a new star \remove{with probability $\pi_t$} \new{with probability $\alpha$} or join existing star $k = 1, \ldots, n_t$ with probability $\pi_{kt}$\remove{.
These probabilities are specified as} \new{, defined as}
\begin{align}
	\pi_{kt} = \frac{M_{k,t}^{\gamma}}{\alpha+ \sum_{j = 1}^{n_t} M_{j,t}^{\gamma}} \text{.}
\label{eq:PAstars}
\end{align}

\remove{$\pi_t = \min \left (1, \alpha \right ) \;\;\; \text{and} \;\;\; \pi_{kt} \propto M_{k,t}^{\gamma}$}

\noindent The generating process is complete when the total mass of formed stars \remove{reaches} \new{first exceeds} $\Mtot$. 
The possible ranges of the three parameters are $\lambda > 0$, $\alpha \in [0,1)$, and $\gamma > 0$. 

\remove{The parameter $\alpha$ defines the probability that entering mass forms a new star in a cluster.}  For the growth component, the model allows for linear ($\gamma = 1$), sublinear ($\gamma < 1$), and superlinear ($\gamma > 1$) behavior; the limiting case of $\gamma \to 0$ gives a uniform attachment model.  Finally, the parameter $\lambda$ acts as a scaling factor which controls the average `coarseness' of masses joining the forming stellar cores. 

\begin{figure} \centering
\begin{subfigure}{0.47\textwidth}
\centering
\includegraphics[width = .95\textwidth]{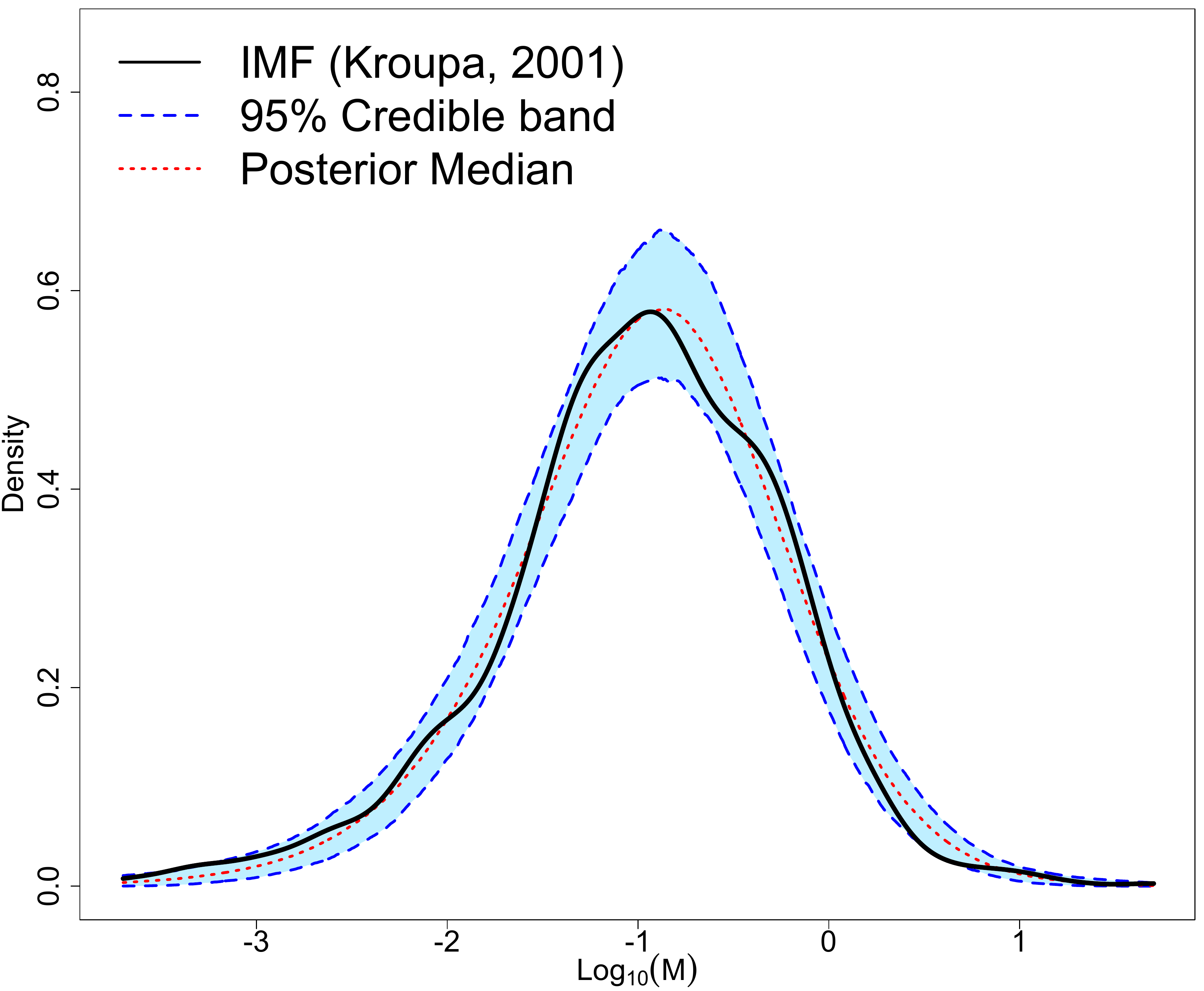} 
\caption{ABC posterior predictive IMF for \cite{kroupa2001} }\label{subfig:kroupa_abc_imf}
\end{subfigure}
\begin{subfigure}{0.47\textwidth}
\centering
\includegraphics[width = .95\textwidth]{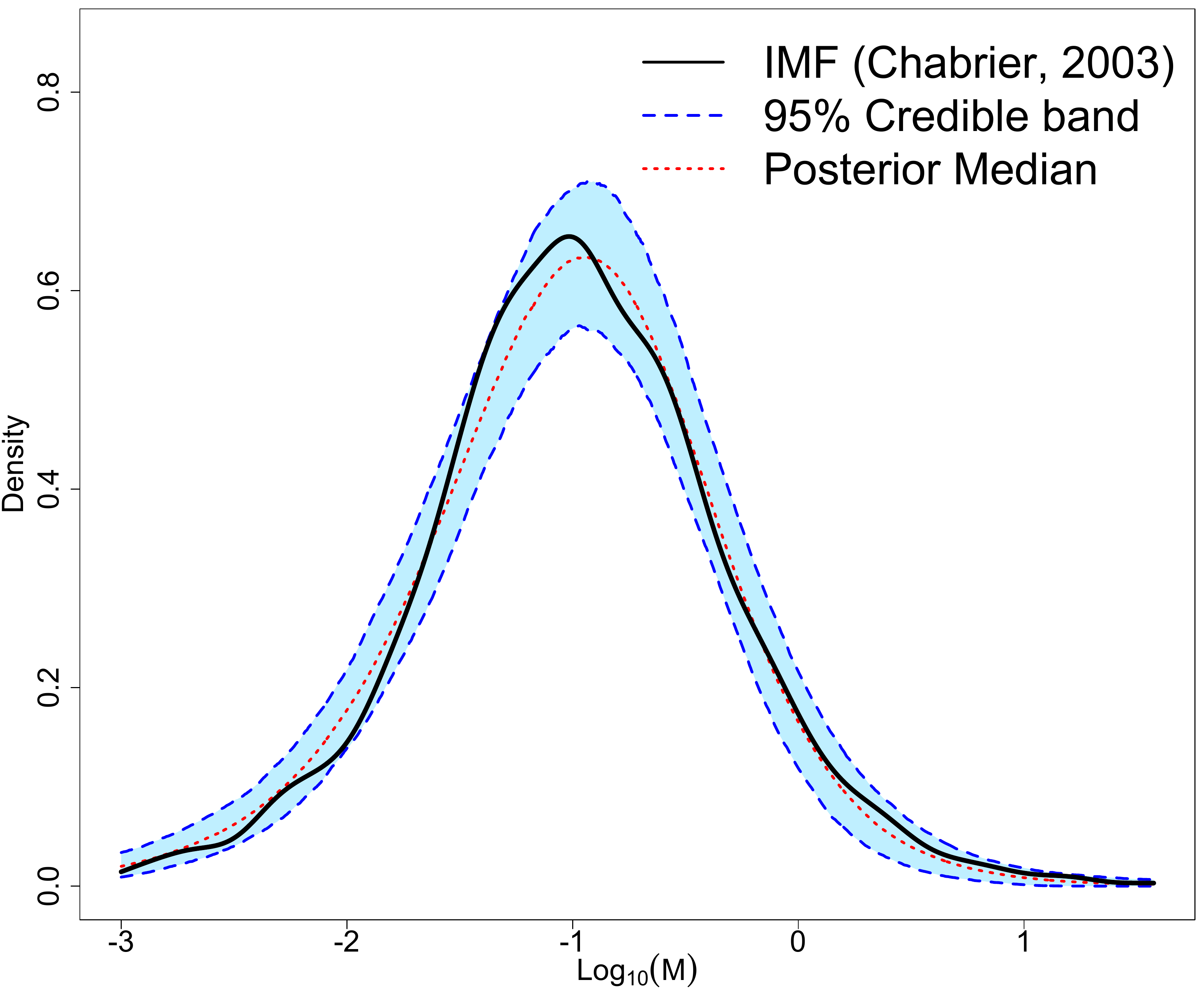} 
\caption{ABC posterior predictive IMF for \cite{Chabrier:2003oq, Chabrier:2003om}}\label{subfig:chabrier_abc_imf}
\end{subfigure}
%
\caption{The solid black lines display the IMF with 1,000 stars simulated from (a) the broken power-law model of \cite{kroupa2001} (see Eq~\eqref{eq:kroupa}) and (b) the log-normal model of \cite{Chabrier:2003oq, Chabrier:2003om}  (see Eq~\eqref{eq:chab}).  The proposed PA ABC model was used with $N = 1,000$ particles, and 95\% point-wise credible bands are displayed (blue, dashed lines) along with the posterior median (red, dotted) for each data set.  The PA model provides flexibility to approximate both existing models.
Both models have similar ABC posterior means for $\alpha$ (0.293 and 0.304, respectively).  However, the ABC posterior means for $\gamma$ are notably different.  The broken power-law model has an ABC posterior mean of 0.889 while the estimate for the log-normal model is 1.050.
	}\label{fig:otherModels}
\end{figure}

\begin{figure}[htbp]
\begin{subfigure}{0.32\textwidth}
\centering
\includegraphics[width = .95\textwidth]{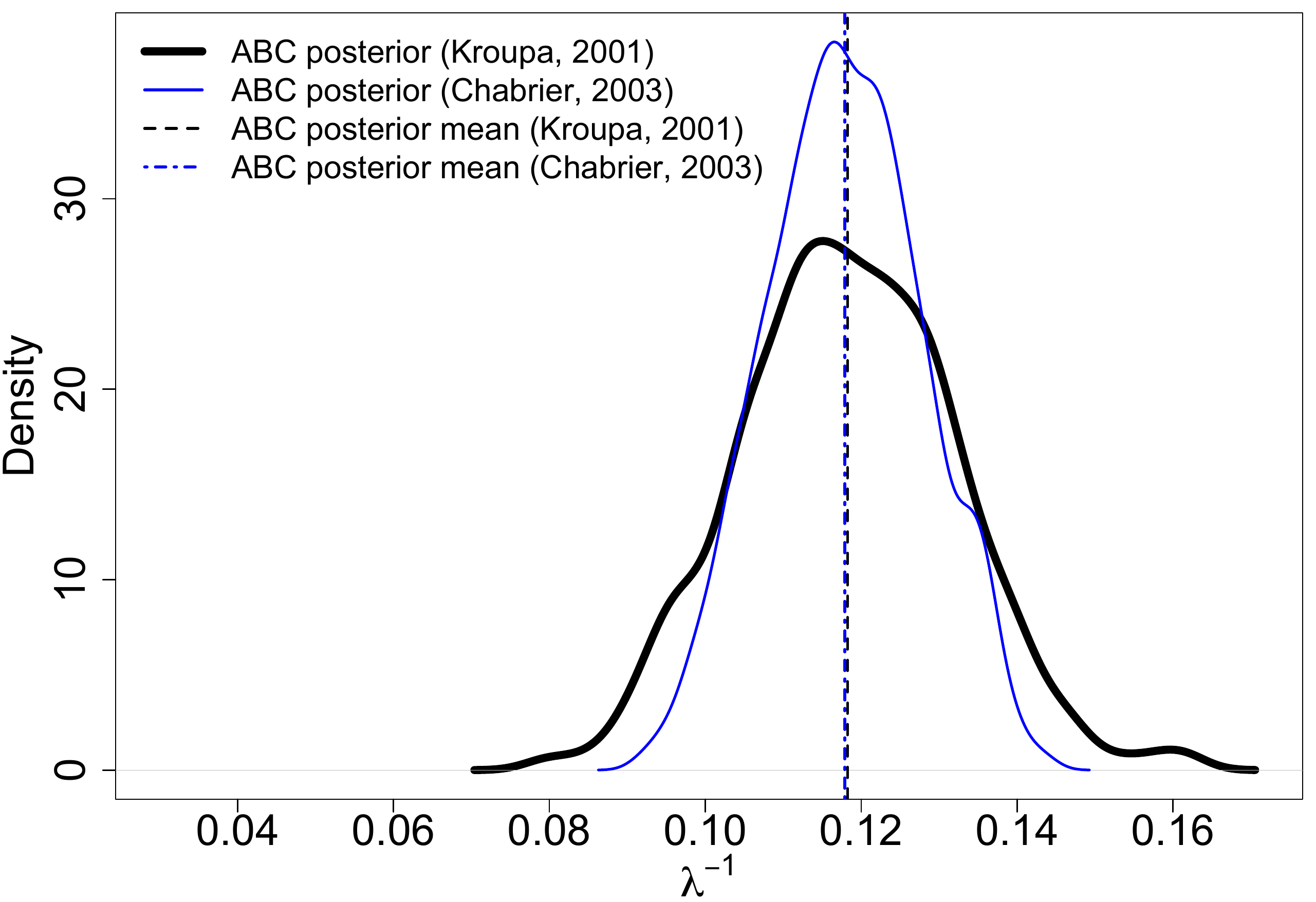} 
\caption{ABC posteriors for $\lambda^{-1}$}\label{subfig:kc_k}
\end{subfigure}
\begin{subfigure}{0.32\textwidth}
\centering
\includegraphics[width = .95\textwidth]{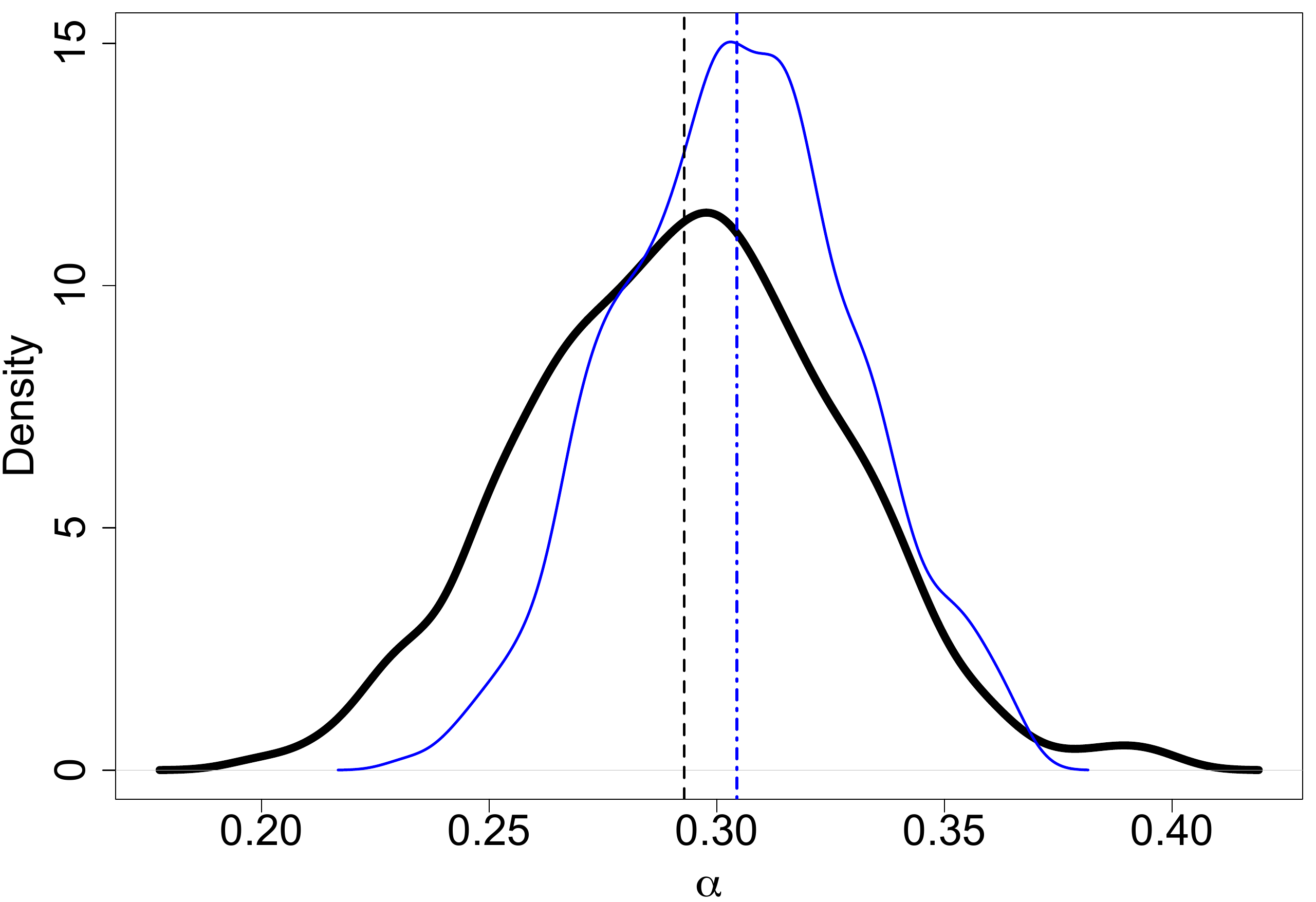} 
\caption{ABC posteriors for $\alpha$}\label{subfig:kc_alpha}
\end{subfigure}
\begin{subfigure}{0.32\textwidth}
\centering
\includegraphics[width = .95\textwidth]{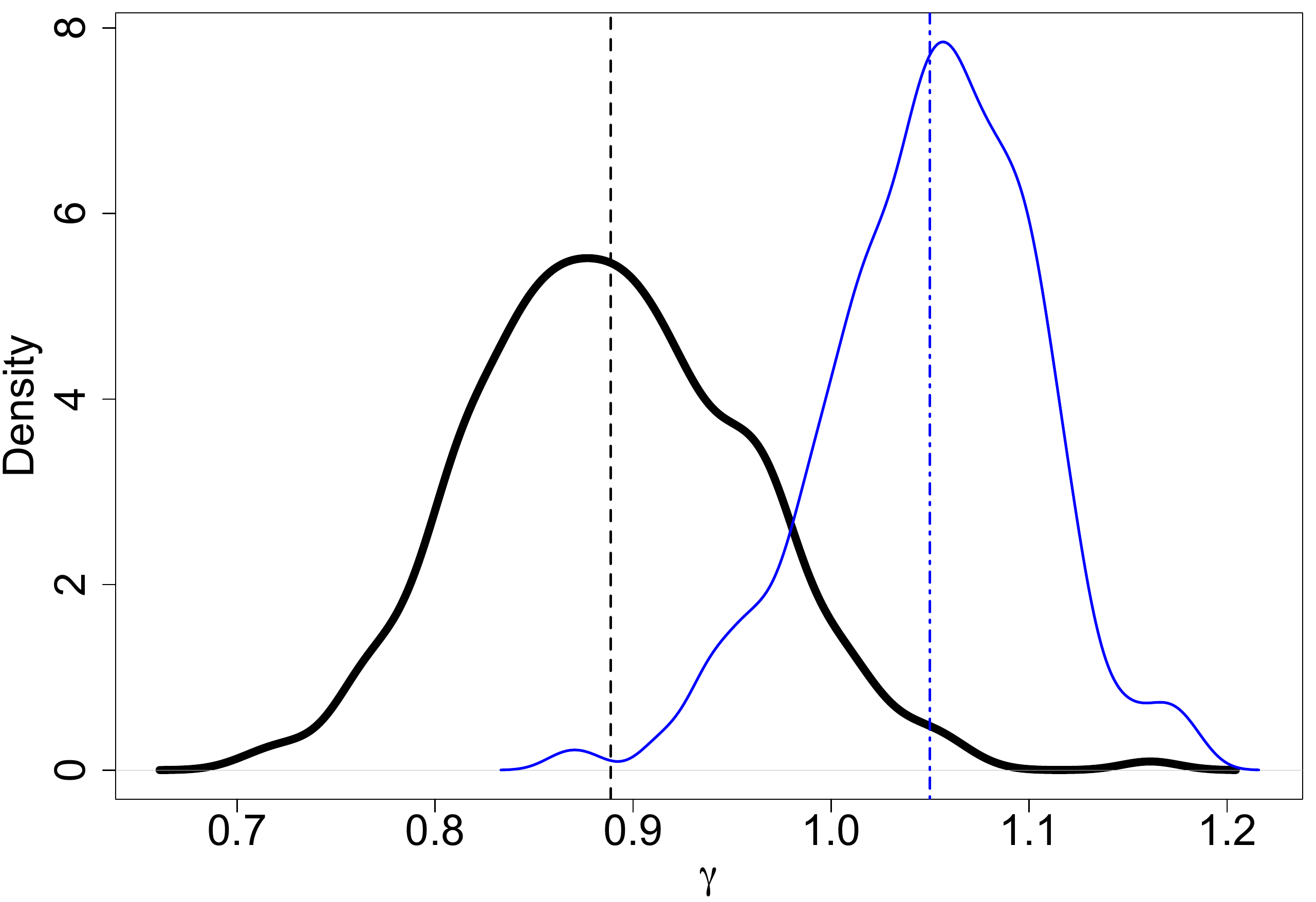} 
\caption{ABC posteriors for $\gamma$}\label{subfig:kc_gamma}
\end{subfigure} \\
\caption{Marginal ABC posteriors for data generated from the broken power-law model of \cite{kroupa2001} (thick black lines)
and for data generated from the log-normal model of \cite{Chabrier:2003oq, Chabrier:2003om} (thin blue lines).  The vertical dashed black lines indicate the ABC posterior mean for the \cite{kroupa2001} model, vertical dashed and dotted blue lines indicate the ABC posterior mean for the \cite{Chabrier:2003oq, Chabrier:2003om} model, and the dotted gray lines indicate the range of the uniform prior for the parameter.  
 }
\label{fig:kroupa_chab_marginals}
\end{figure}

The proposed PA mass generation model offers considerable flexibility to approximate existing IMF models in the literature.  To illustrate the generality of the proposed model, IMF realizations were drawn assuming the \cite{kroupa2001} 
broken power-law model as the true model, defined as
\begin{equation}
f(m) \propto \left\{
  \begin{array}{lr}
   m^{-0.3}, &  m \leq 0.08 \\
   k_1 \cdot m^{-1.3}, &  0.08 < m \leq 0.5 \\
   k_2 \cdot m^{-2.3}, &  m > 0.5,
  \end{array}
\right. \label{eq:kroupa}
\end{equation}
and the \cite{Chabrier:2003oq, Chabrier:2003om} log-normal model, defined as
\begin{equation}
f(m) \propto \left\{
  \begin{array}{lr}
    \frac{0.158}{m} \times \exp \left ( -\frac{(\log_{10}(m) - \log_{10}(0.079))^2}{2(0.69)^2}\right), &  m \leq 1\\
   k_3 \cdot m^{-2.3}, &  m > 1,
  \end{array}
\right. \label{eq:chab}
\end{equation}
where constants $k_1$, $k_2$, and $k_3$ are defined to make the densities continuous.
Our proposed PA ABC procedure was then used for inference and Figure~\ref{fig:otherModels} displays the resulting posterior predictive IMFs.  The proposed model captures the general shape of the true model.  Figures~\ref{subfig:kc_k} - \ref{subfig:kc_gamma} display ABC marginal posteriors for the broken power-law model of \cite{kroupa2001} and the log-normal model of  \cite{Chabrier:2003oq, Chabrier:2003om}.  Both the broken power-law and log-normal models have similar ABC posterior means for $\alpha$ (0.293 and 0.304, respectively).  However, the ABC posterior means for $\gamma$ are notably different.  The broken power-law model has an ABC posterior mean of 0.889 while the estimate for the log-normal model is 1.050.  Since the  \cite{kroupa2001} and \cite{Chabrier:2003oq, Chabrier:2003om} models use the same power-law slope for masses greater than 0.5 $\Msun$ and 1 $\Msun$, respectively, this suggests that the differences in $\gamma$ are due to differences in the shape of the lower-mass end of the IMF.  The proposed model offers an approach for discriminating these models.

\subsection{Initial Mass Function to the Observed Mass Function} \label{sec:obs_challenges}
The PA model describes the formation of a star cluster at initial formation.  \remove{However, we are not generally able to observe the star cluster after initial formation due to observational uncertainties, measurement uncertainties, and aging and dynamical evolution of the cluster.  A cluster's present-day observed MF is the observed distribution of the stellar masses of a particular cluster.}
\new{In practice, we are not generally able to observe the star cluster after the initial formation because significant time is likely to have passed.  When observation of a cluster occurs, the initial cluster will have changed due to aging and dynamical evolution of the cluster.  Additionally, even if observation of the initial cluster was possible, there are observational and measurement uncertainties that would limit our capacity to get a perfect representation of the initial cluster.  The actual observed cluster is referred to as the \emph{present-day observed MF}, which describes the observed distribution of the stellar masses of a particular cluster.}

Observation limitations can be easily incorporated into the ABC framework.  For simplicity, we adopt a ``linear ramp" completeness function describing the probability of observing a star of mass $m$:
\begin{align}
	\Proba(\text{observing a star} \mid \emph{m}) = \begin{cases} 0 \text{,} & m < \Cmin\\
								\frac{m - \Cmin}{\Cmax - \Cmin}\text{,} & m \in [\Cmin, \Cmax] \\
								1 \text{,} & m > \Cmax \text{.}
								\end{cases} \label{eq:ramp}
\end{align}
We assume that the values $\Cmin$ and $\Cmax$ are known, though we note that selecting an appropriate completeness function is a difficult process which requires quantification from the observational astronomers for each set of data.
Different models for the completeness function could also be considered, including those which allow for spatially-varying observational completeness.  A benefit of ABC is the ease at which a new completeness function can be incorporated -- it amounts to a simple change in the forward model.  

Due to measurement error and practical limitations in translating luminosities into masses, the masses of stars are not perfectly known. 
This uncertainty can be incorporated in different ways; following \cite{weisz13}, we assume that the inferred mass of a star $m_i$ is related to its true mass $M_i$ via 
\begin{align}
	\log m_i = \log M_i + \sigma_i \eta_i \text{,}
	\label{eq:masserror}
\end{align}
where $\eta_i$ is a standard normal random variable, and $\sigma_i$ is known measurement error.   
The model for mass uncertainties in \eqref{eq:masserror} is simple and could be extended to account for other sources of uncertainty (e.g. redshift).

As noted previously, the lifecycle of a star depends on certain characteristics such as mass.  In the proposed algorithm, stars generated in a cluster are aged using a simple truncation of the largest masses.  
That is, the distribution of stellar masses for a star cluster of age $\tau$ Myr is given by
	\begin{align}
	f_M(m \mid \theta, \tau) \propto f_M(m \mid \theta) \indic \{M \leq \tau^{-1/3} \times 10^{4/3}\} \text{,}
	\label{eq:age}
	\end{align}
\remove{$f_M(m \mid \theta, \tau) = f_M(m \mid \theta) \indic \{M \leq \tau^{-1/3} \times 10^{4/3}\}$}

\noindent corresponding to stellar lifetimes of (10$^4/M^{3}$) Myr, where $M$ is the mass of the star \citep{hansen2004, Chaisson:2011}\new{, and $f_M(m \mid \theta)$ represents some specified IMF model}. 
More sophisticated models that account for effects such as binary stars and stellar wind mass loss can be inserted into this framework.  
\section{Simulation study} \label{sec:sim}

We propose an ABC framework to make inferences on the IMF given a cluster's present-day observed MF.\footnote{Code for running the proposed ABC-IMF algorithm is available at \url{https://github.com/JessiCisewskiKehe/ABC_for_StellarIMF}.}  \new{Details about the proposed ABC method, including the algorithm, are presented in Appendix~\ref{methodSec:abc}.  In this section we first consider a simulation study where the data are generated from the proposed forward model with observational effects, and then we consider data from an astrophysical simulation  \citep{Bate2012, Bate2014}}.

\subsection{Simulated data with observational effects} \label{sec:sim_obs}

\remove{Next we} \new{We first} consider a suite of simulations which incorporate
aging, completeness, and measurement error, in order to analyze these effects on the resulting inference on the IMF.  
The same IMF was used throughout the simulations, with
$\lambda^{-1} = 0.25$, $\alpha = 0.3$, $\gamma = 1$, and $\Mtot = 1000$ \remove{(setting number 2 from Table)}, but we vary the range of the linear ramp completeness function of Equation~\eqref{eq:ramp}; $\Cmin$ is fixed at 0.08 $\Msun$ and $\Cmax \in \{0.10, 0.25, 0.5, 0.75, 1\}$ where low values of $\Cmax$ result in fewer stars removed from the IMF and, hence, a larger number of stars in the MF.  
All five sets of MFs are aged 30 Myr and have log-normal measurement error with $\sigma = 0.25$.  
The observational effects, including the differing completeness function upper bounds, resulted in MF's with 800 (70.1\% of IMF stars), 659 (57.7\%), 488 (42.7\%), 415 (36.3\%), and 352 (30.8\%) stars for $\Cmax = 0.10, 0.25, 0.5, 0.75, 1$, respectively, compared to the original IMF with 1142 stars.

\begin{figure}[htbp]
   \centering
\begin{subfigure}{0.48\textwidth}
\centering
\includegraphics[width = \textwidth]{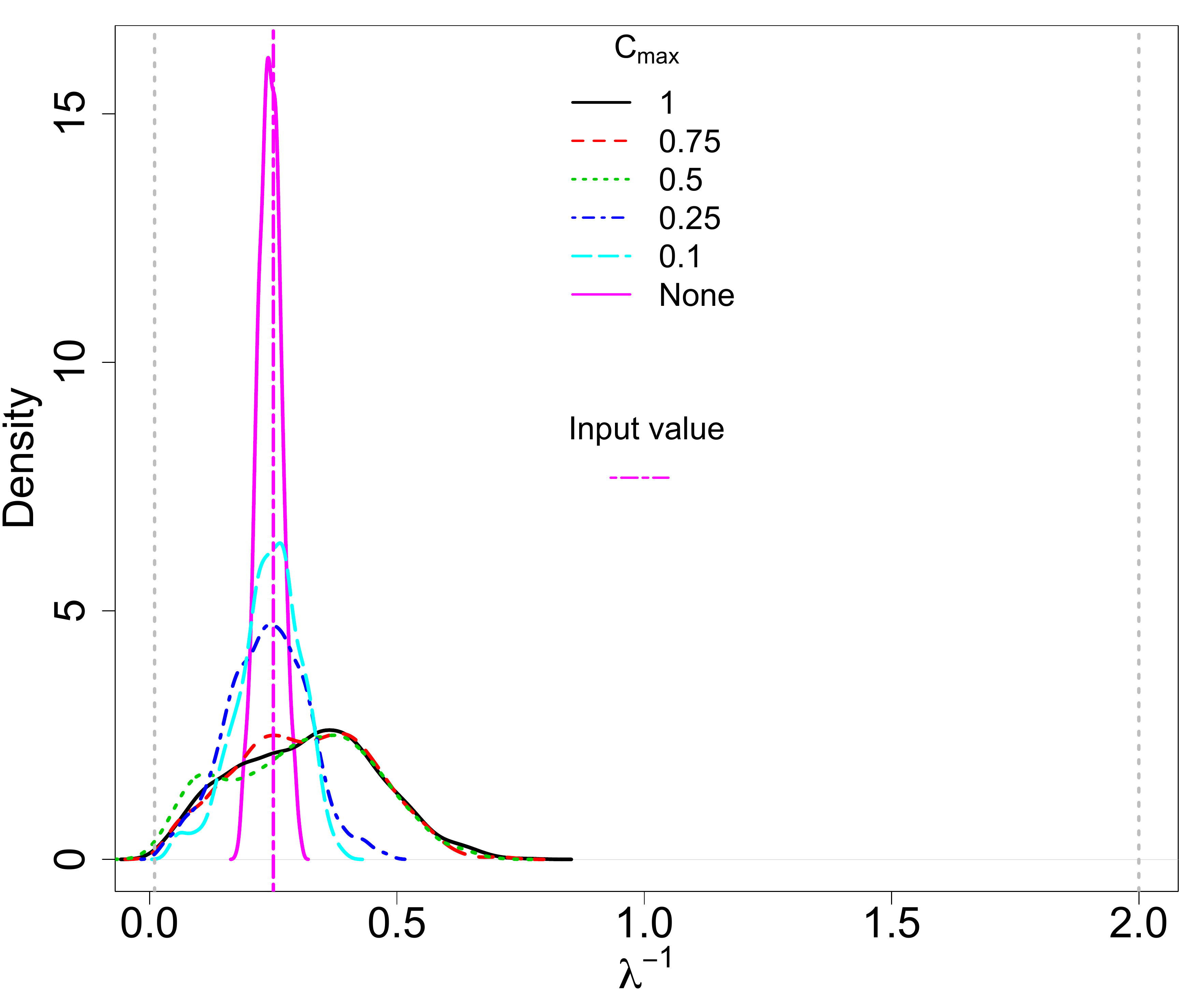} 
\caption{ABC marginal posterior for $\lambda^{-1}$}\label{subfig:marg_k_obs}
\end{subfigure}
\begin{subfigure}{0.48\textwidth}
\centering
\includegraphics[width = \textwidth]{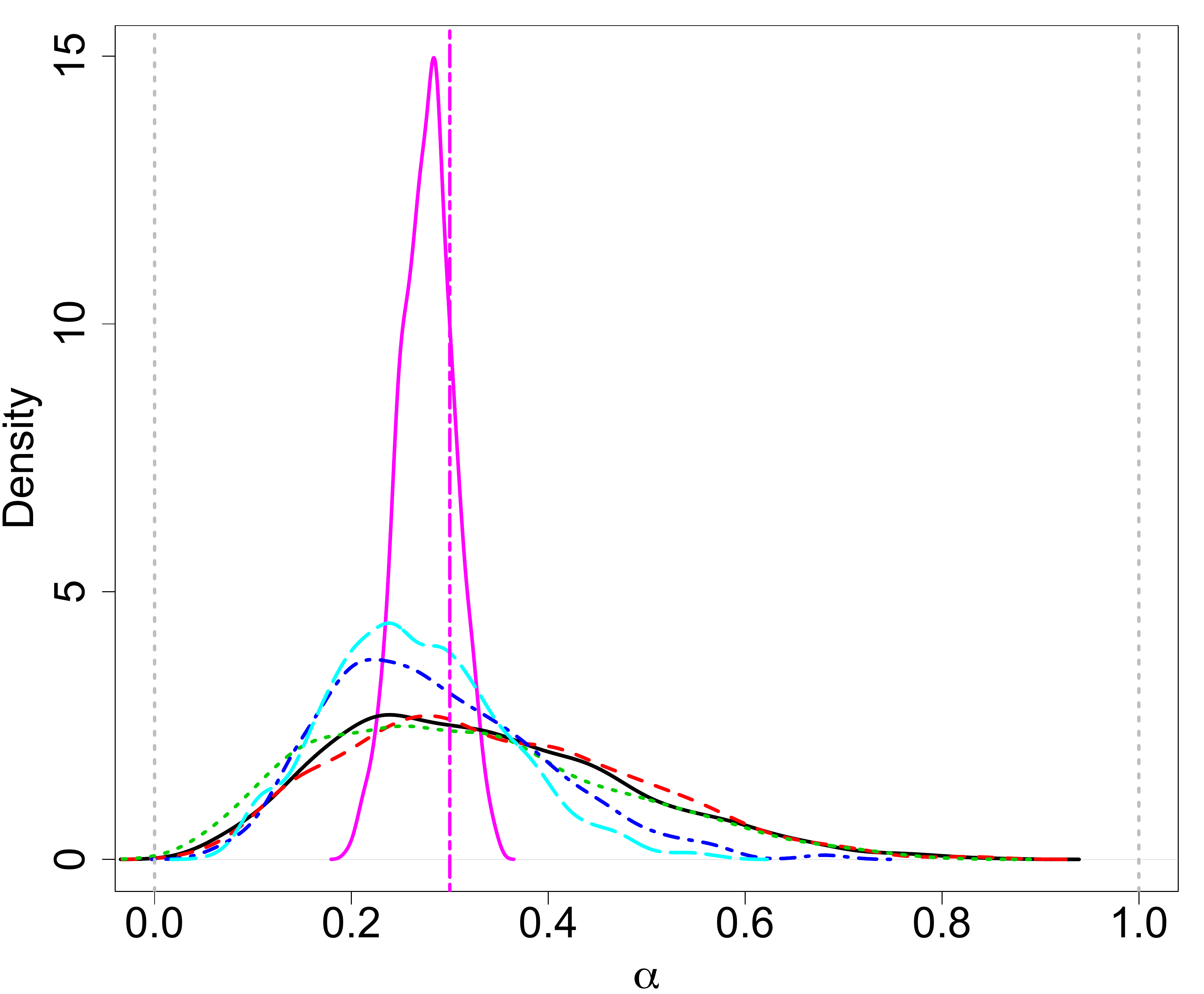} 
\caption{ABC marginal posterior for  $\alpha$}\label{subfig:marg_alpha_obs} 
\end{subfigure} \\
\begin{subfigure}{0.48\textwidth}
\centering
\includegraphics[width = \textwidth]{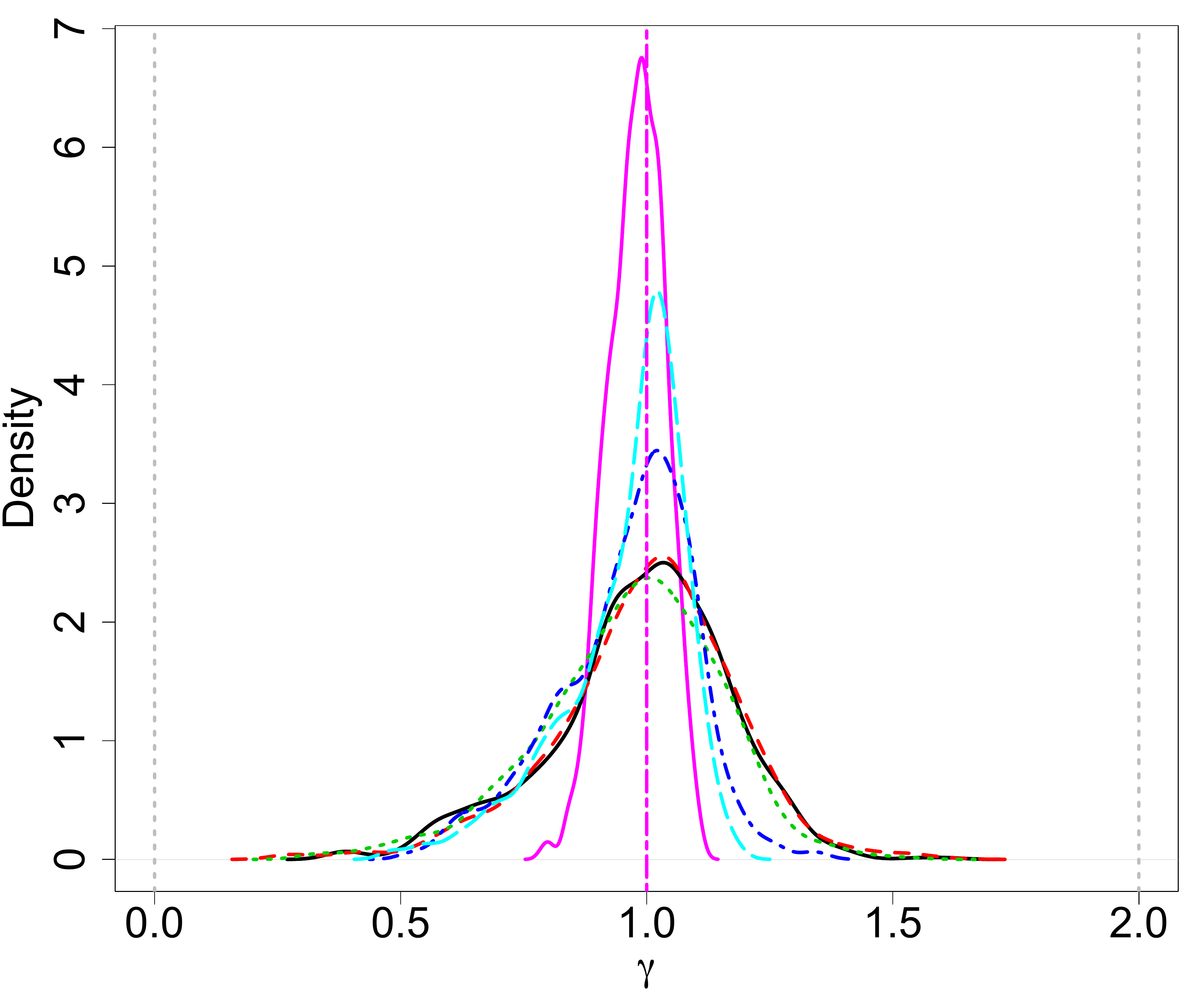} 
\caption{ABC marginal posterior for $\gamma$}\label{subfig:marg_gamma_obs}
\end{subfigure}
\begin{subfigure}{0.48\textwidth}
\centering
\includegraphics[width = \textwidth]{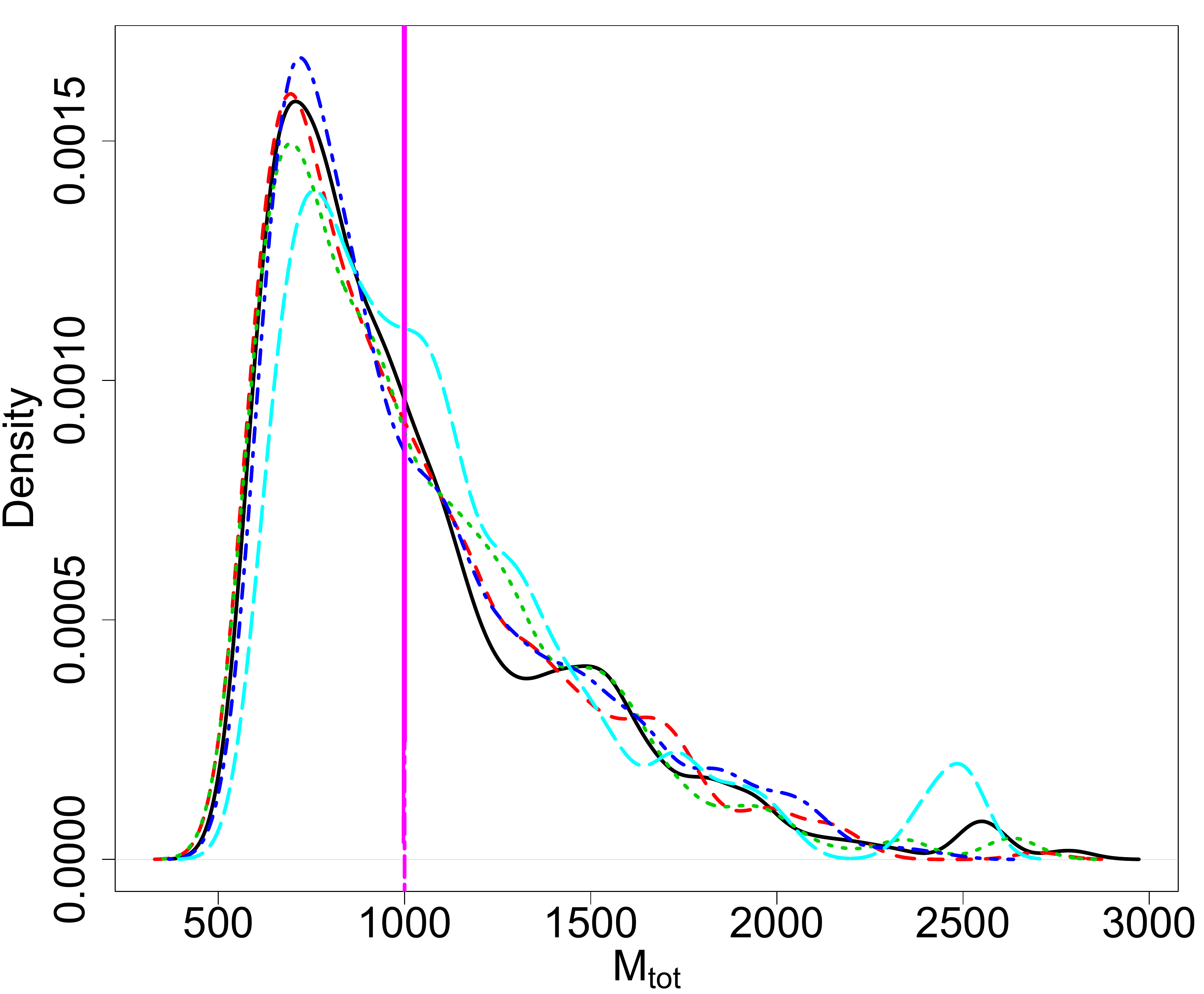} 
\caption{ABC marginal posterior for  $\Mtot$}\label{subfig:marg_mtot_obs}
\end{subfigure}
 \caption{Marginal ABC posteriors for the simulation setting of Section~\ref{sec:sim_obs}.  The different color and types of lines indicate the differing upper limits of the linear ramp completeness function of Equation~\eqref{eq:ramp}, $\Cmax$, corresponding to the weighted kernel density estimates of the marginal ABC posteriors for (a) $\lambda^{-1}$, (b) $\alpha$, (c) $\gamma$, and (d) $\Mtot$.  
The lower limit, $\Cmin$, is fixed at 0.08 $\Msun$.  
All five datasets started with the same IMF using $\lambda^{-1} = 0.25$, $\alpha = 0.3$, $\gamma = 1$, and $\Mtot = 1000$, were aged 30 Myr, and had log-normal measurement error applied with $\sigma = 0.25$.
The solid magenta line is the ABC marginal posterior using an identical IMF, but with no observational effects applied \remove{(i.e. the same ABC marginal posteriors as the red dashed lines in Figure)}, and is included for comparison; note that the vertical axis of (d) does not extend to the full range of this ABC marginal posterior for $\Mtot$.
The vertical dotted gray lines indicate the range of the priors for (a), (b), and (c).
   }
   \label{fig:abc_pa_posterior_obs}
\end{figure}

\begin{figure}[htbp]
   \centering
\begin{subfigure}{0.32\textwidth}
\centering
\includegraphics[width = \textwidth]{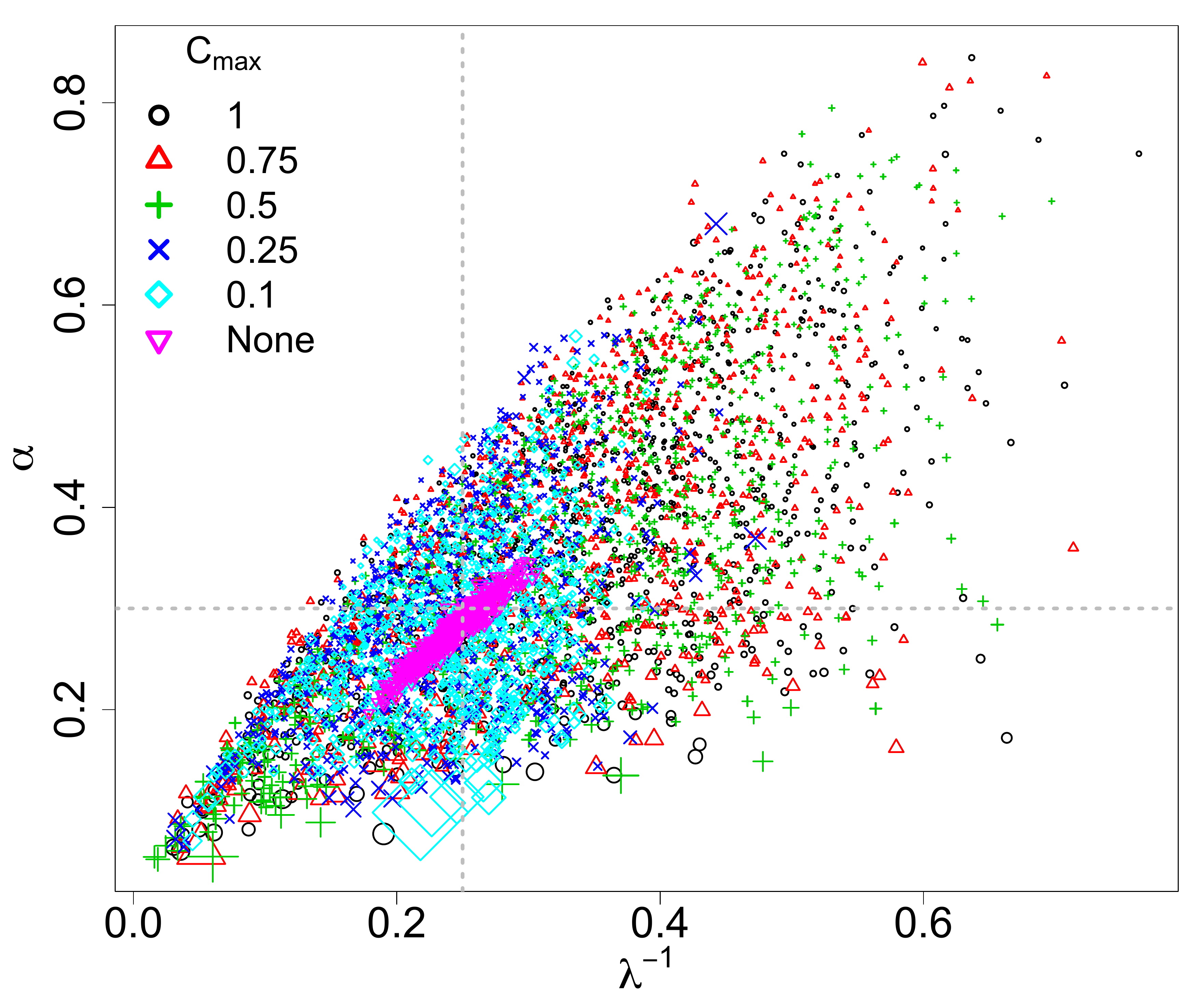} 
\caption{ABC Joint $(\lambda^{-1}, \alpha)$}\label{subfig:joint_alpha_k_obs}
\end{subfigure}
\begin{subfigure}{0.32\textwidth}
\centering
\includegraphics[width = \textwidth]{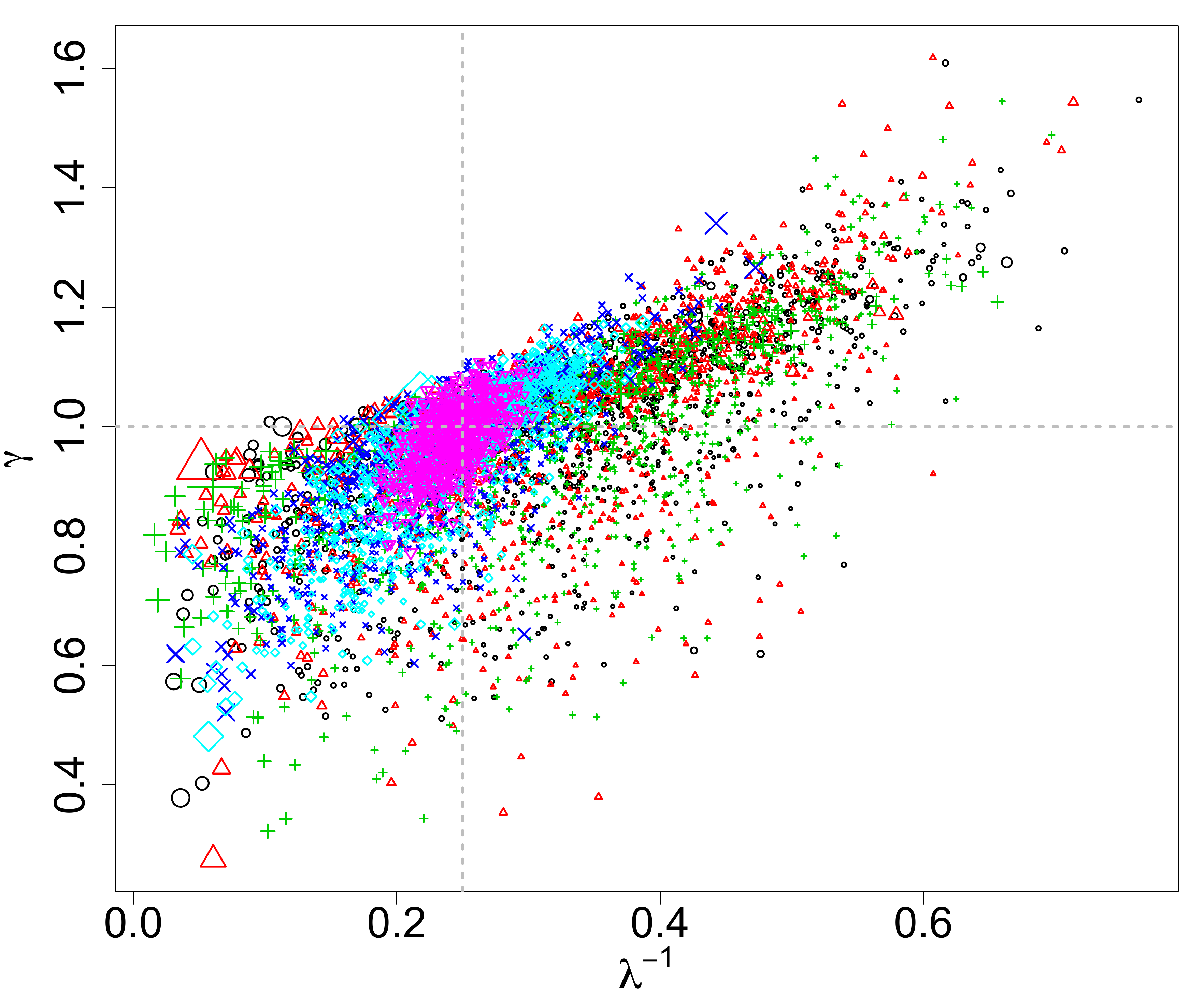} 
\caption{ABC Joint $(\lambda^{-1}, \gamma)$}\label{subfig:joint_gamma_k_obs}
\end{subfigure}
\begin{subfigure}{0.32\textwidth}
\centering
\includegraphics[width = \textwidth]{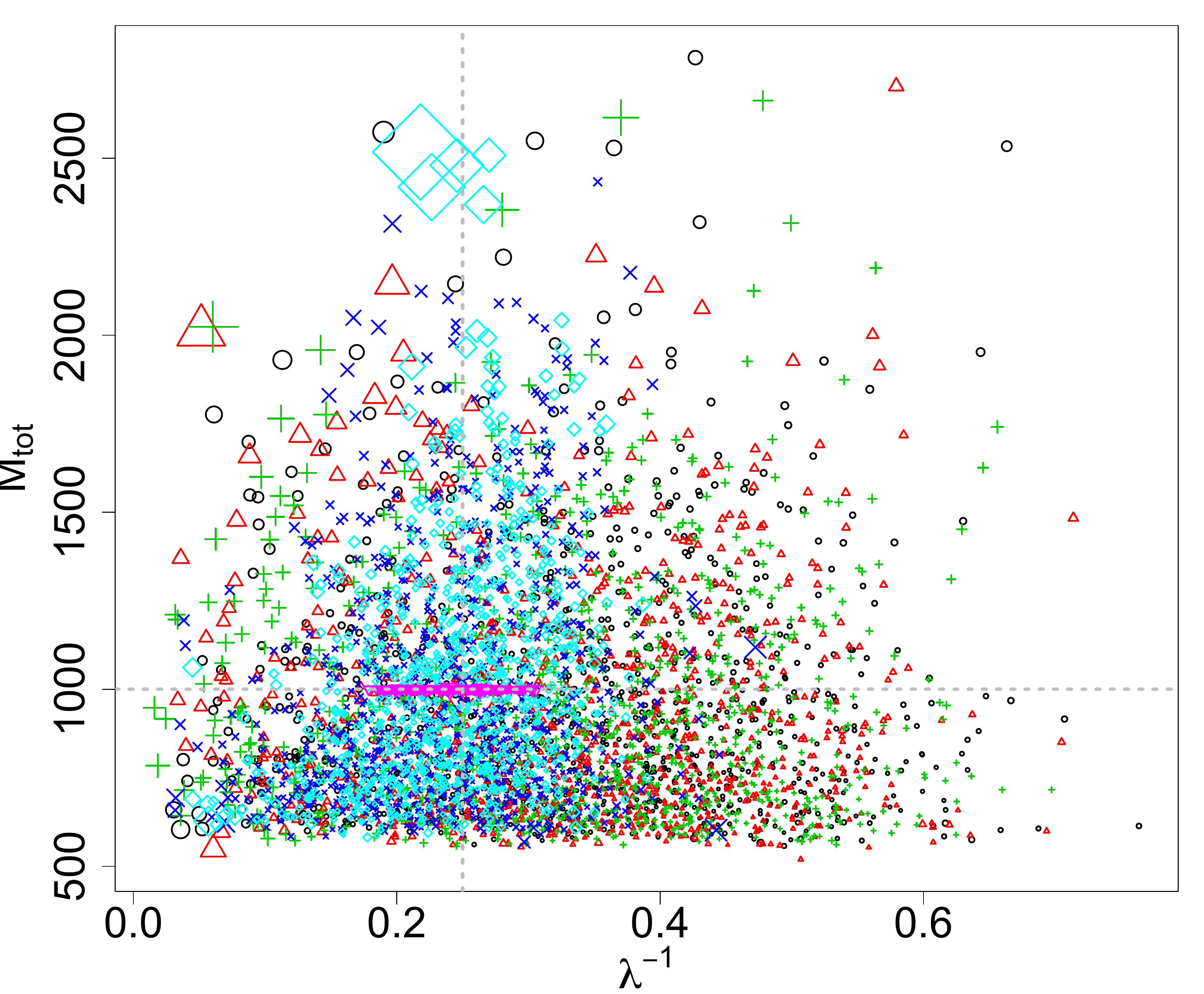} 
\caption{ABC Joint $(\lambda^{-1}, \Mtot)$}\label{subfig:joint_mtot_k_obs}
\end{subfigure} \\
\begin{subfigure}{0.32\textwidth}
\centering
\includegraphics[width = \textwidth]{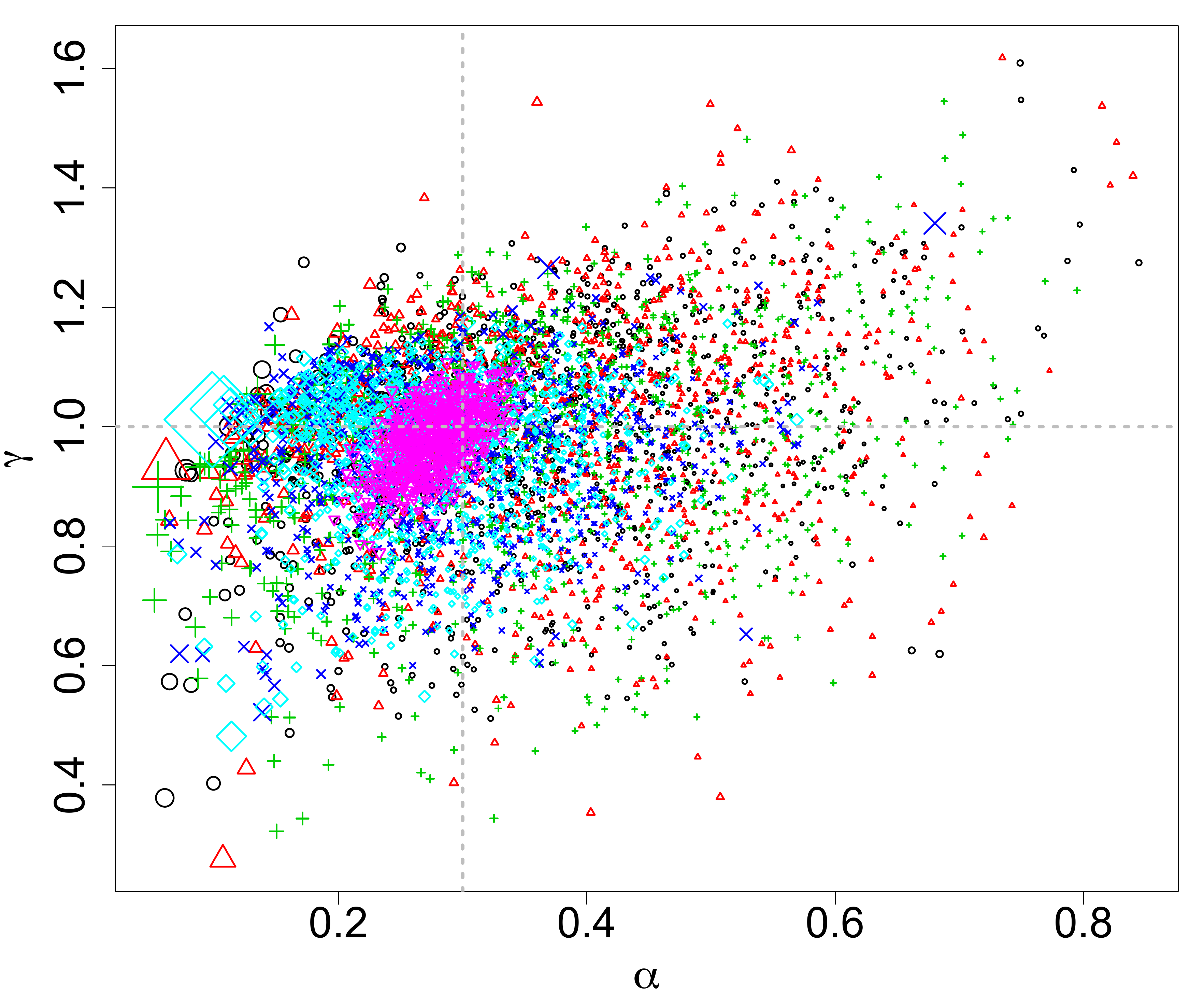} 
\caption{ABC Joint $(\alpha, \gamma)$}\label{subfig:joint_gamma_alpha_obs}
\end{subfigure}
\begin{subfigure}{0.32\textwidth}
\centering
\includegraphics[width = \textwidth]{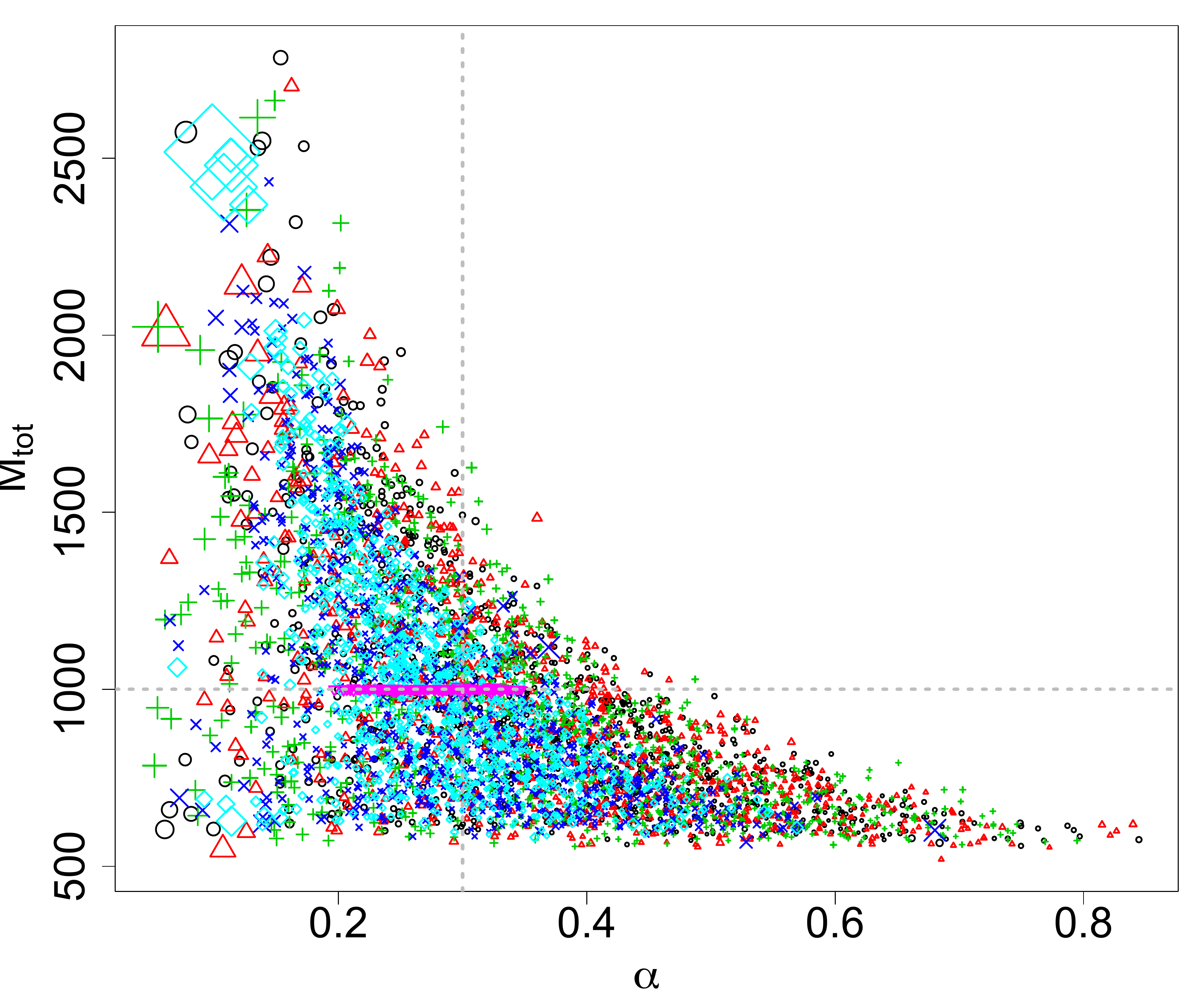} 
\caption{ABC Joint $(\alpha, \Mtot)$}\label{subfig:joint_mtot_alpha_obs}
\end{subfigure}
\begin{subfigure}{0.32\textwidth}
\centering
\includegraphics[width = \textwidth]{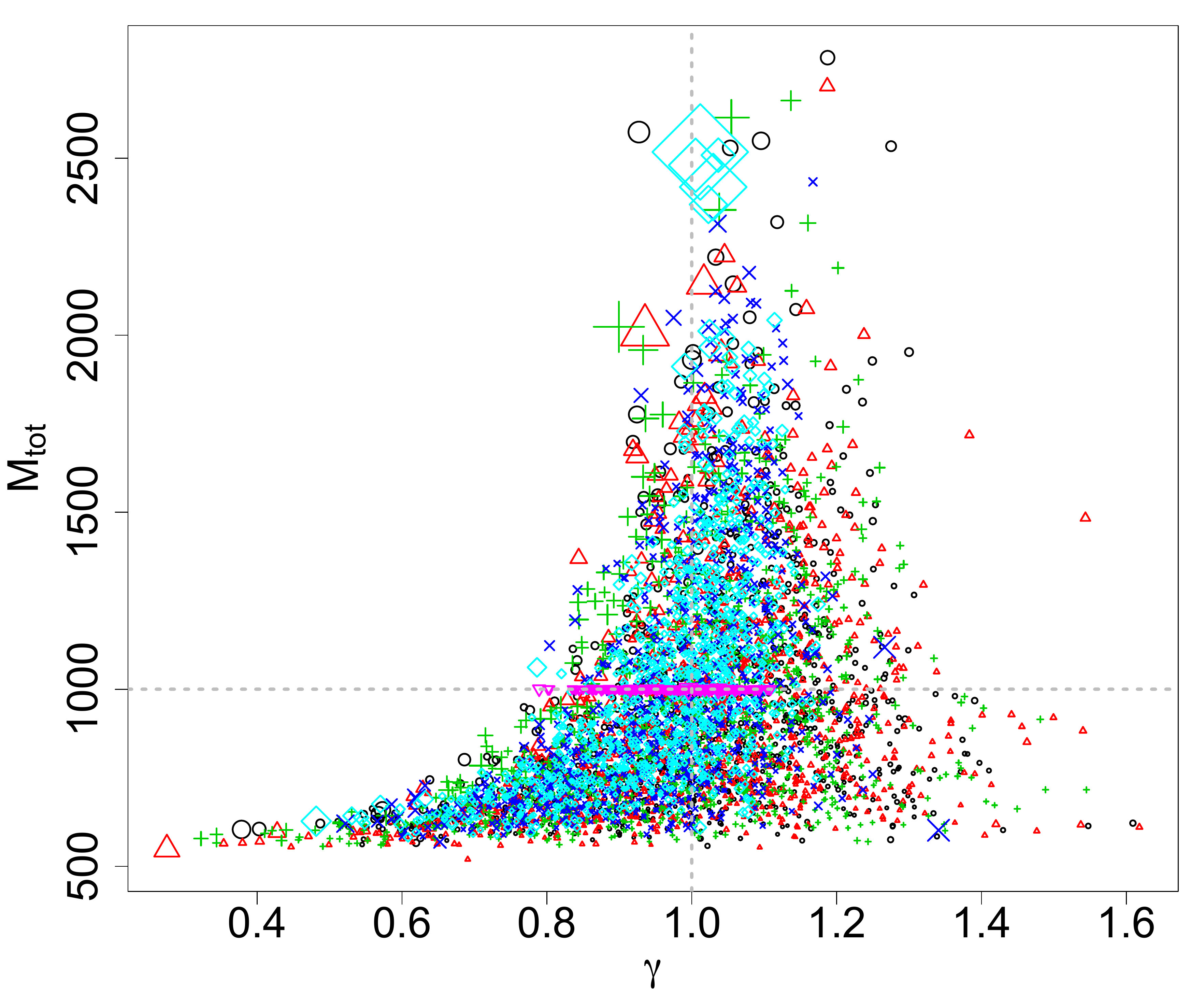} 
\caption{ABC Joint $(\alpha, \gamma)$}\label{subfig:joint_mtot_gamma_obs}
\end{subfigure} \\
\caption{Pairwise ABC posterior particles samples of $(\lambda^{-1}, \alpha, \gamma, \Mtot)$ for the simulation settings of Section~\ref{sec:sim_obs}.  The different color and types of points indicate the $\Cmax$ values of the linear ramp completeness function, and the size of the plot symbol is scaled with the particle weight.  
The lower limit, $\Cmin$, is fixed at 0.08 $\Msun$.  
All five datasets started with the same IMF using $\lambda^{-1} = 0.25$, $\alpha = 0.3$, $\gamma = 1$, and $\Mtot = 1000$, were aged 30 Myr, and had log-normal measurement error applied with $\sigma = 0.25$.
The magenta upside-down triangles are the ABC marginal posterior using an identical IMF, but with no observational effects applied \remove{(i.e. the same ABC joint posteriors as the red triangles in Figure)}, and is included for comparison.
}
\label{fig:abc_pa_joints_obs}
\end{figure}

We are interested in the differences among the ABC posteriors and predictive IMFs among the varying $\Cmax$ values.  
The marginal ABC posteriors are displayed in Figure~\ref{fig:abc_pa_posterior_obs}, which also includes the analogous ABC marginal posteriors without observational effects \remove{from Figure}.  
Except for the marginal posteriors of $\Mtot$, for $\Cmax = 0.1 \text{ and } 0.25$, the posteriors get broader, which is expected because larger $\Cmax$ results in fewer observations and greater uncertainty.  
However, the marginal posteriors for $\Cmax = 0.5, 0.75$, and $1$ are quite similar. The marginal posteriors of $\Mtot$ in Figure~\ref{subfig:marg_mtot_obs} are all similar and significantly broader than the case without observational effects.  Hence, the observational effects appear to have a profound impact on inference for $\Mtot$.  The pairwise joint ABC posteriors are displayed in Figure~\ref{fig:abc_pa_joints_obs} as a reference, and seem to follow the same general patterns noted for the marginals (i.e., they are broader as $\Cmax$ increases).

\begin{figure}[htbp]
\centering
\includegraphics[width=.85\textwidth]{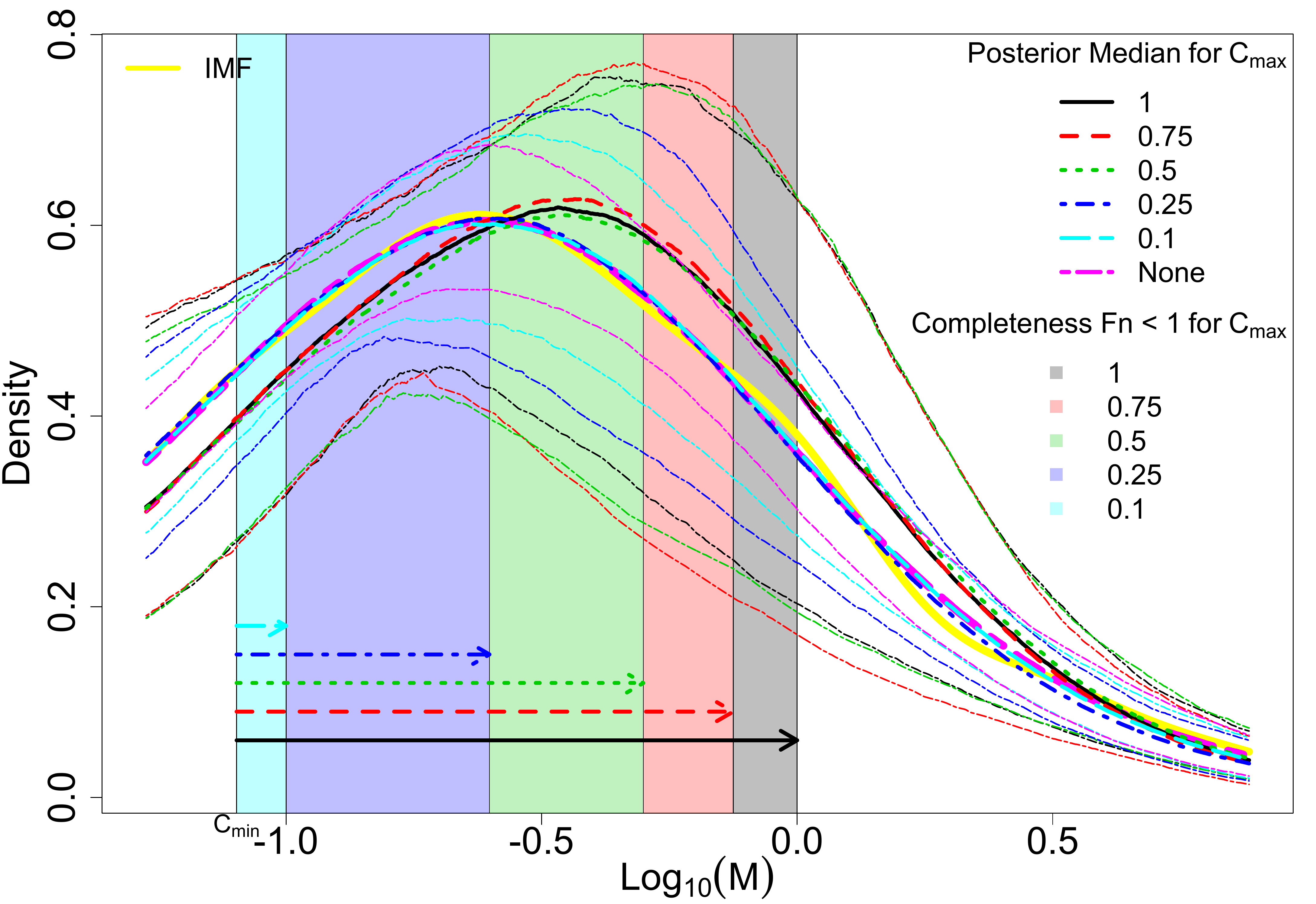}
 \caption{Posterior predictive IMF for the simulation settings of Section~\ref{sec:sim_obs}. 
The thick yellow line is the true IMF, the thicker lines of varying line type and color are the ABC posterior predictive median IMF for the different values of $\Cmax$, 
the thinner lines of the same type and color define a point-wise 95\% credible band  for the $\Cmax$ with the matching color, and
the shaded regions are the different ranges of completeness (all starting at $\Cmin = 0.08 \Msun$ indicated by the left end of the arrows).
The solid magenta line is the ABC posterior predictive IMF using an identical IMF, but with no observational effects applied \remove{(i.e. the same ABC posterior predictive IMFs as the red dashed line in Figure)}, and is included for comparison.
The posterior predictive IMFs are based off of 1000 independent draws from the final ABC posteriors.  
} \label{fig:abc_pa_pred_obs}
\end{figure}

Finally, the posterior predictive IMFs are combined into a single plot displayed in Figure~\ref{fig:abc_pa_pred_obs}.  As in the previous section, the posterior predictive IMFs are the pointwise medians of 1000 independent draws from the ABC posteriors of Figure~\ref{fig:abc_pa_posterior_obs}.  Also included in the figures are 95\% credible bands based on the 2.5 and 97.5 percentiles of the 1000 posterior draws.  The true IMF is plotted as a thick yellow line and the corresponding ABC posterior predictive IMF without observational effects \remove{(the red dashed line of Figure)} is also displayed.
The posterior predictive IMFs for $\Cmax = 0.1$ and $0.25$ overlap well with the true IMF and the posterior predictive IMF without observational effects, but with wider 95\% credible bands.
The posterior predictive IMFs for $\Cmax = 0.5, 0.75,$ and $1$ have similar shapes and 95\% credible bands.  
Their posterior predictive IMFs peak at a higher mass than the others.  These differences are not surprising given that there are far fewer stars below, for example, $-0.5 \log_{10}(\Msun)$:  
46, 61, and 96 stars for $\Cmax = 1, 0.75, \text{ and } 0.5$ compared to 225 and 368 stars for $\Cmax = 0.25, \text{ and } 0.1$, respectively, and 685 stars in that range for the original IMF.

The conclusion drawn from these simulations is that the completeness function affects the resulting inference -- when more stars are missing from the original IMF due to the completeness function, the resulting ABC posteriors tend to be broader to reflect the increased uncertainty. 

\subsection{Astrophysical simulation data} \label{sec:bate}

\begin{figure}[htbp]
\centering
\includegraphics[width = .5\textwidth]{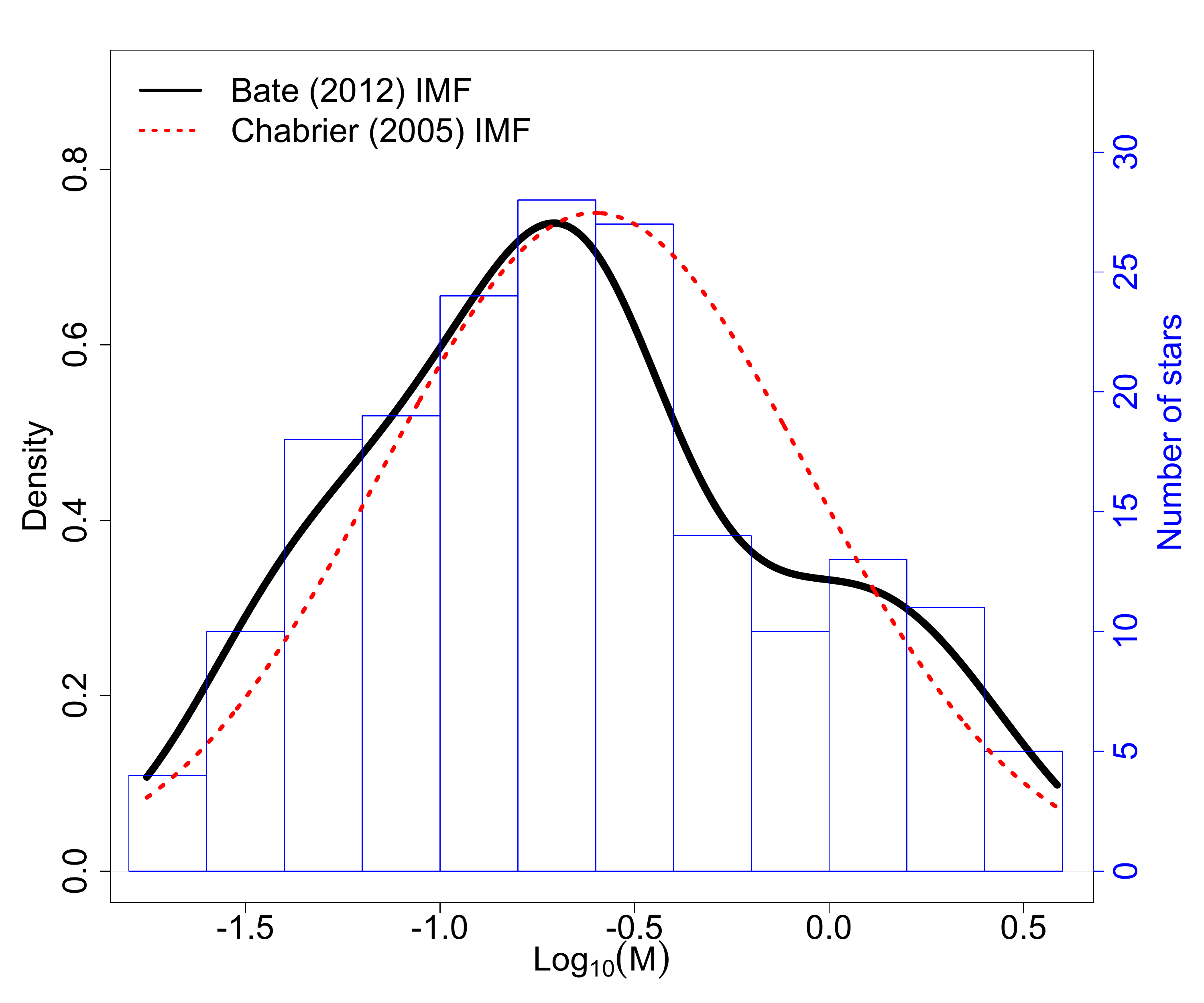} 
\caption{Astrophysical simulation IMF.  The black curve displays the IMF of the 183 stars and brown dwarfs simulated from \cite{Bate2012} and the red dotted curve is the IMF of \cite{chabrier2005}.  The right axis provides the number of stars (and brown dwarfs) for the histogram (plotted in blue).
   }
   \label{fig:bate1}
\end{figure}

Next we consider a star cluster generated from the radiation hydrodynamical simulation presented in \cite{Bate2012} and published in \cite{Bate2014}.\footnote{The astrophysical data is available at \url{https://ore.exeter.ac.uk/repository/handle/10871/14881}}  
This simulation resulted in 183 stars and brown dwarfs with a total mass of the resulting objects of about 88.68 $\Msun$ formed from a 500 $\Msun$ molecular cloud of uniform density.  
Understanding that simulations are only an approximation of reality, this astrophysical simulation was implemented to include realistic physics of star cluster formation such as radiative feedback. 
The technical details of the simulation are beyond the scope of this work, but can be found in \cite{Bate2012}.
Figure~\ref{fig:bate1} displays the resulting IMF as a density and histogram.

In \cite{Bate2012}, validation of the simulated cluster was carried out by comparing its IMF with the model of \cite{chabrier2005}, and was not able to statistically differentiate them using a Kolmogorov-Smirnov test.  The \cite{chabrier2005} IMF is displayed in Figure~\ref{fig:bate1} as a comparison to the simulation data.  While the general shape does appear to match well, the \cite{Bate2012} data has a small second mode around $1\Msun$.  The \cite{Bate2012} data seems to have more objects on the lower mass end and fewer between 0.5 and 1$\Msun$ than expected with the \cite{chabrier2005} IMF model.
Additionally, because the shape of the low-mass end of the IMF is not well-constrained observationally,  \cite{Bate2012} compares the ratio of number of brown dwarfs to number of stars with masses $<1\Msun$ and finds acceptable agreement with observations. 
\cite{Bate2012} also carryout an analysis of the mechanism(s) behind the shape of the IMF.  
They found that larger objects have had longer accretion times, while lower mass objects tended to have a dynamical encounter that result in the accretion terminating; hence there ended up being, as \cite{Bate2012} described, a ``competition between accretion and dynamical encounters.''  This competition for material seems consistent with the ideas underlying the proposed PA model.

\begin{figure}[htbp]
\begin{subfigure}{0.32\textwidth}
\centering
\includegraphics[width = .95\textwidth]{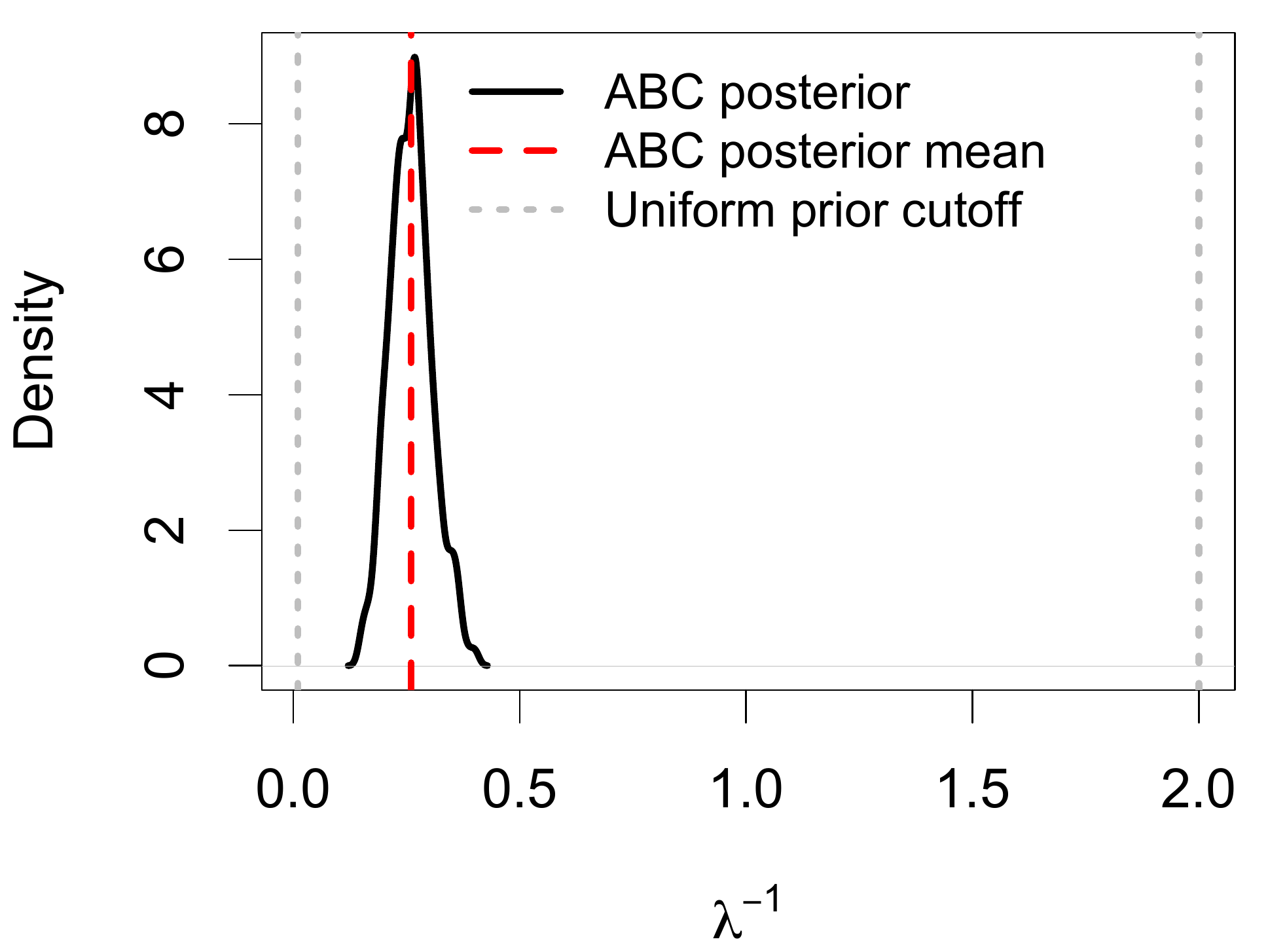} 
\caption{ABC posterior for $\lambda^{-1}$}\label{subfig:bate_k}
\end{subfigure}
\begin{subfigure}{0.32\textwidth}
\centering
\includegraphics[width = .95\textwidth]{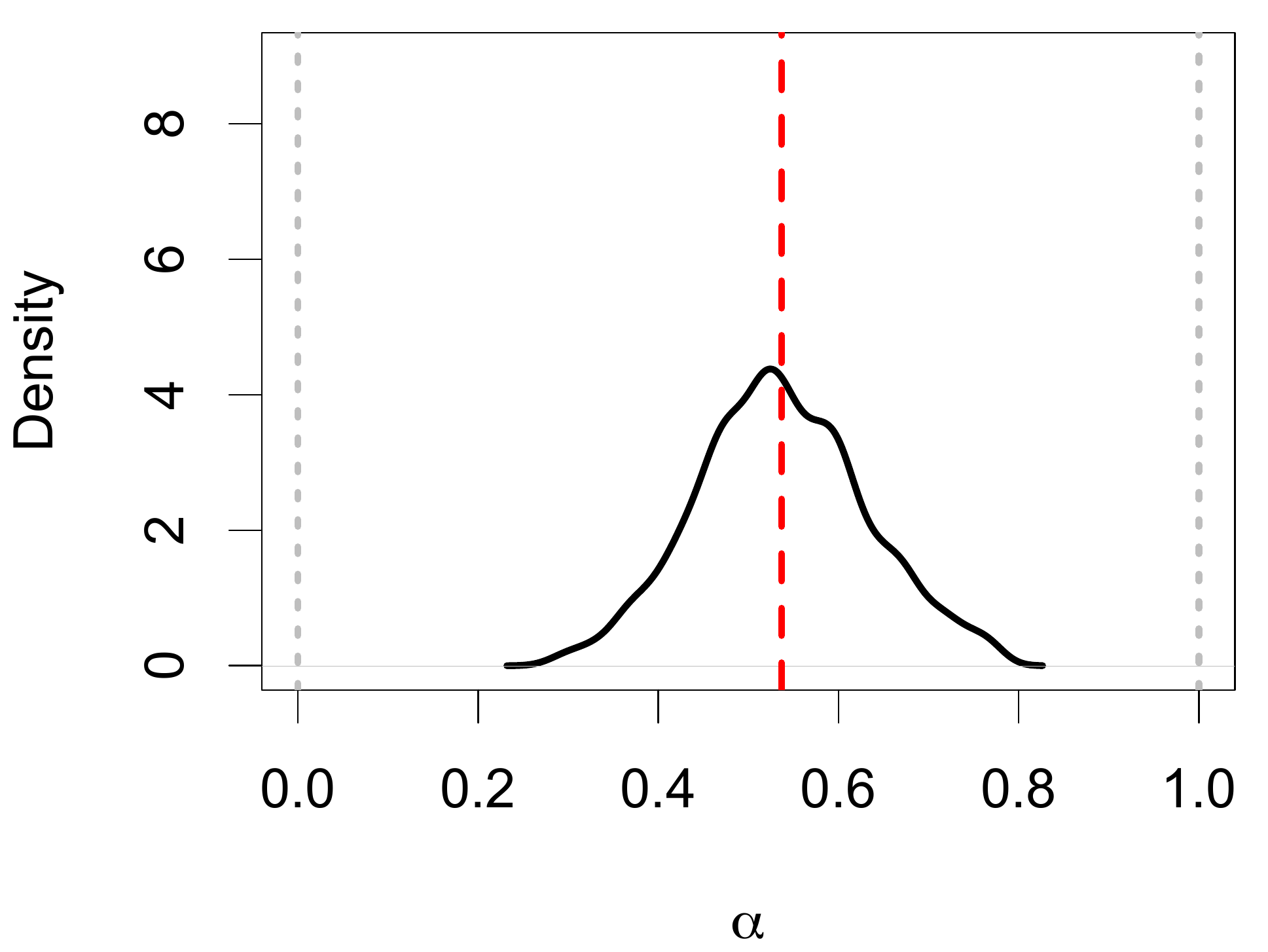} 
\caption{ABC posterior for $\alpha$}\label{subfig:bate_alpha}
\end{subfigure}
\begin{subfigure}{0.32\textwidth}
\centering
\includegraphics[width = .95\textwidth]{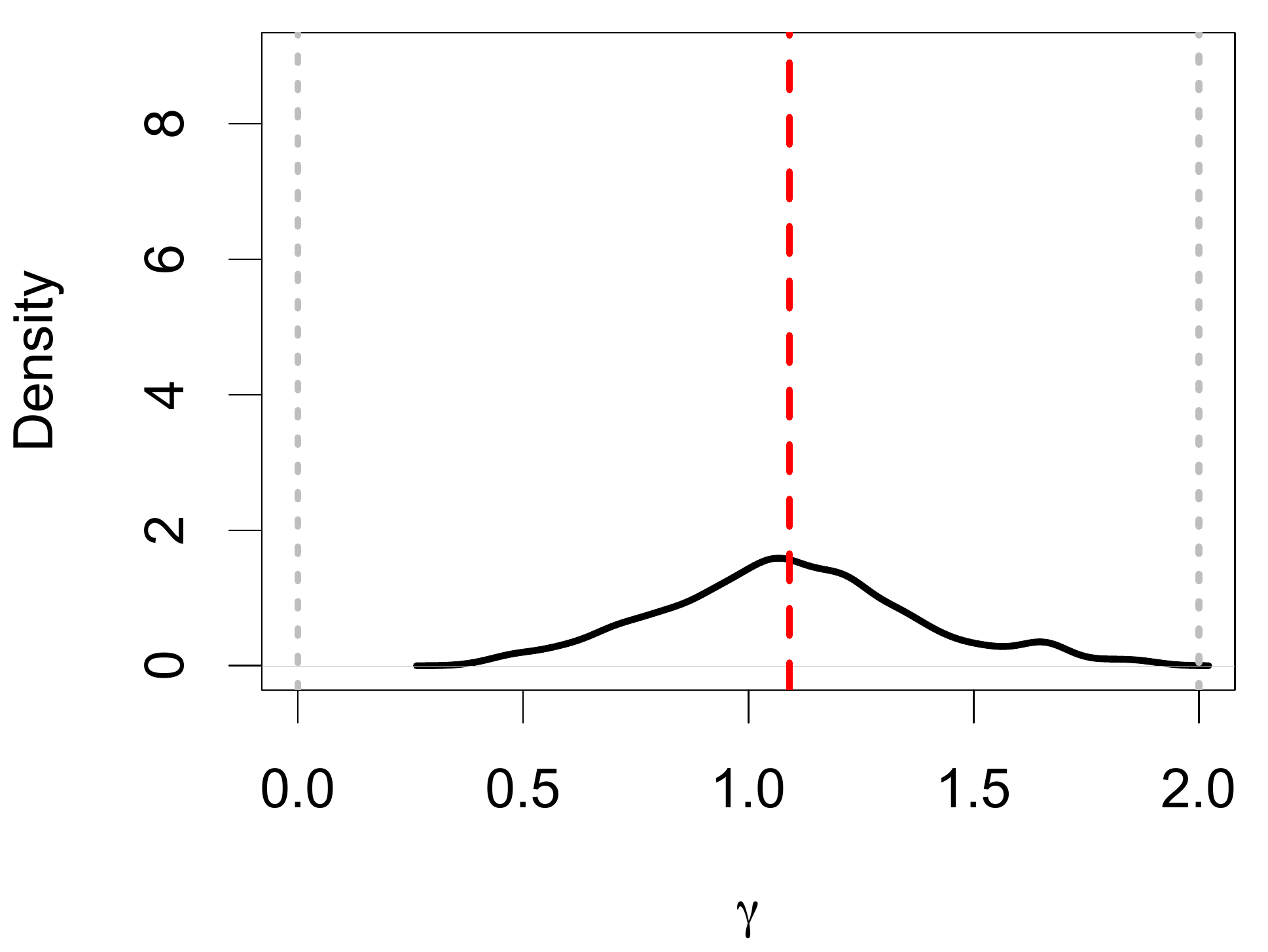} 
\caption{ABC posterior for $\gamma$}\label{subfig:bate_gamma}
\end{subfigure}
\caption{Marginal ABC posteriors for astrophysical simulation data from \cite{Bate2012}.  The vertical dashed red lines indicate the ABC posterior mean, and the dotted gray lines indicate the range of the uniform prior for the parameter.  
 }
   \label{fig:abc_bate_posterior}
\end{figure}

\begin{figure}[htbp]
   \centering
\begin{subfigure}{0.32\textwidth}
\centering
\includegraphics[width = \textwidth]{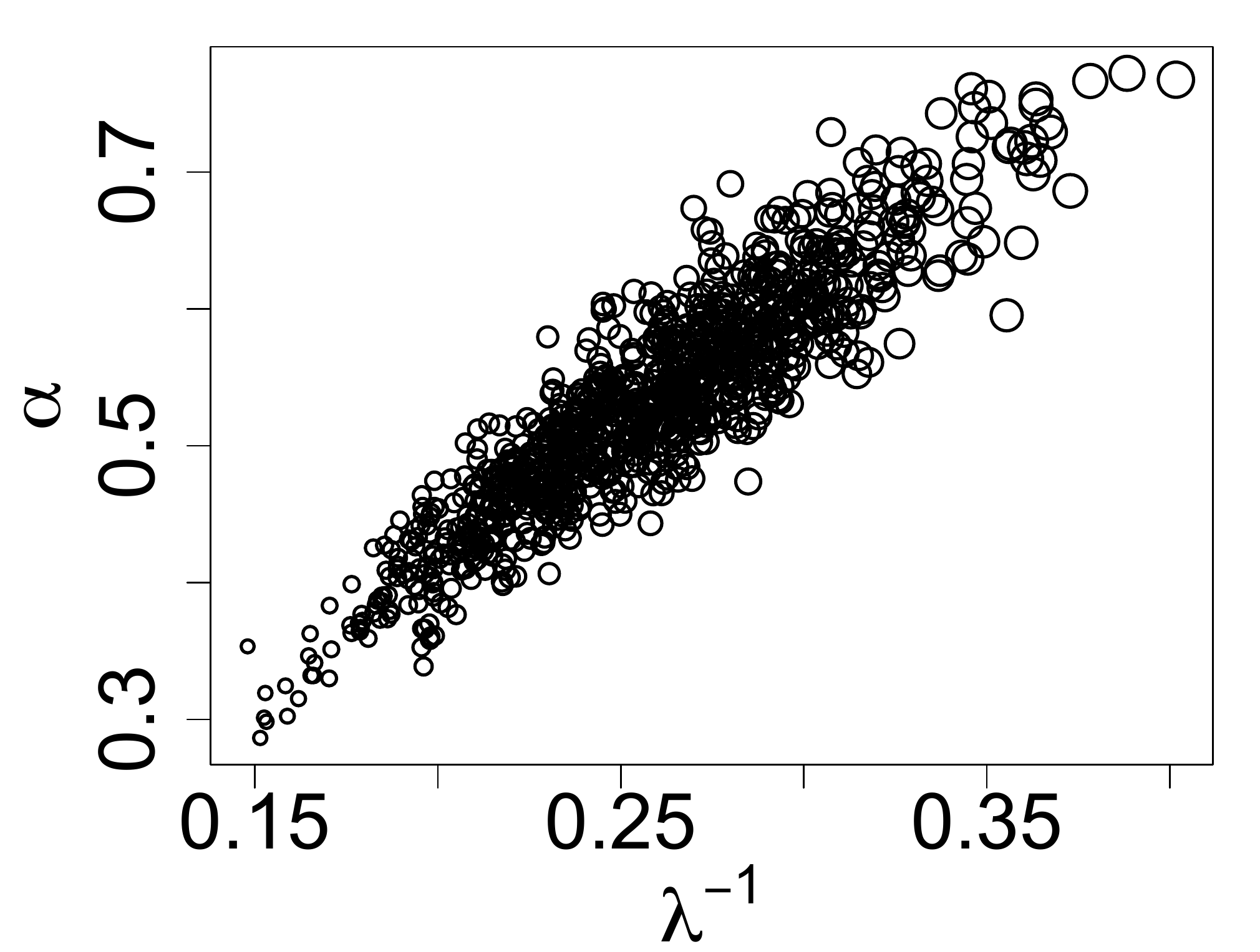} 
\caption{ABC Joint $(\lambda^{-1}, \alpha)$}\label{subfig:joint_alpha_k_bate}
\end{subfigure}
\begin{subfigure}{0.32\textwidth}
\centering
\includegraphics[width = \textwidth]{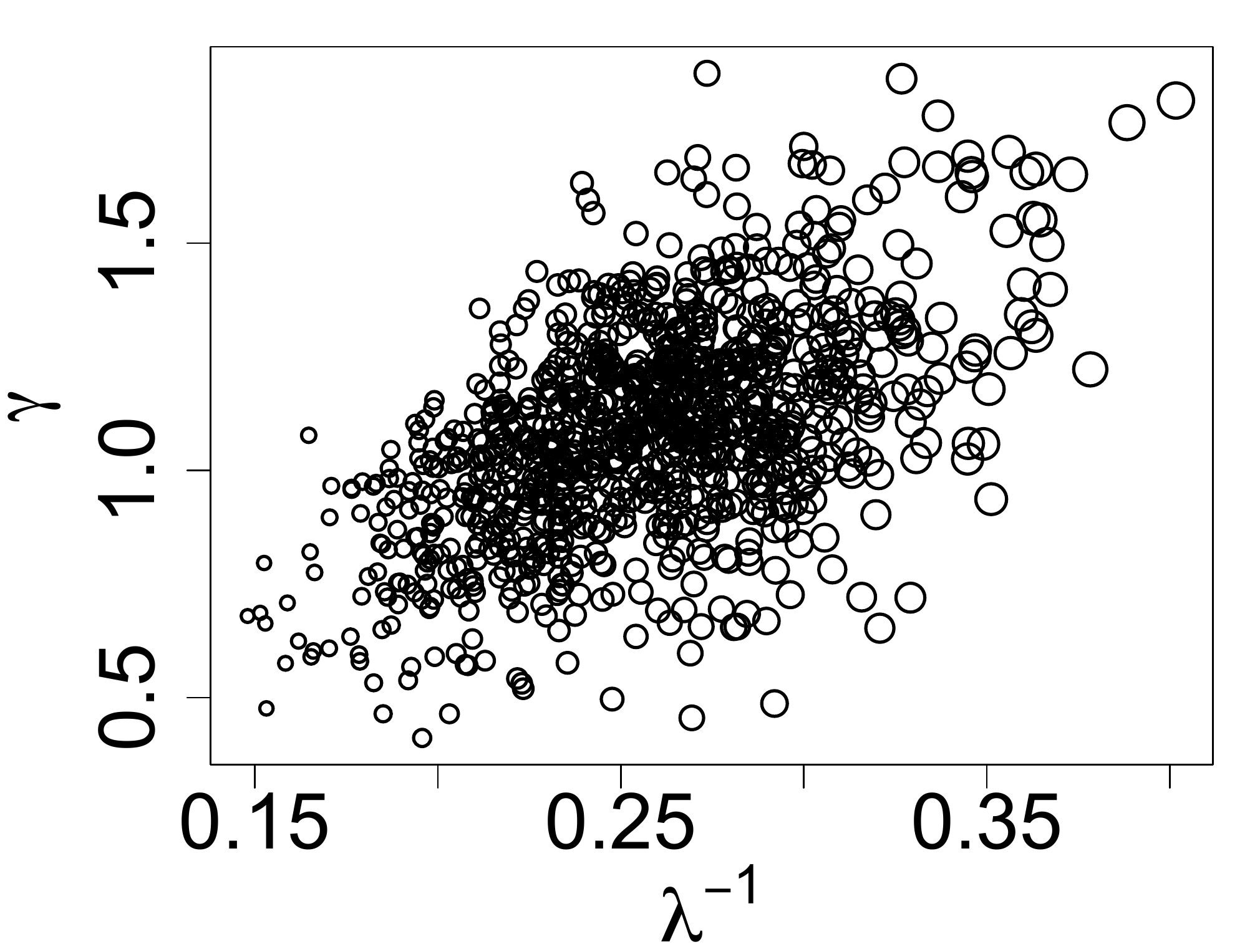} 
\caption{ABC Joint $(\lambda^{-1}, \gamma)$}\label{subfig:joint_gamma_k_bate}
\end{subfigure}
\begin{subfigure}{0.32\textwidth}
\centering
\includegraphics[width = \textwidth]{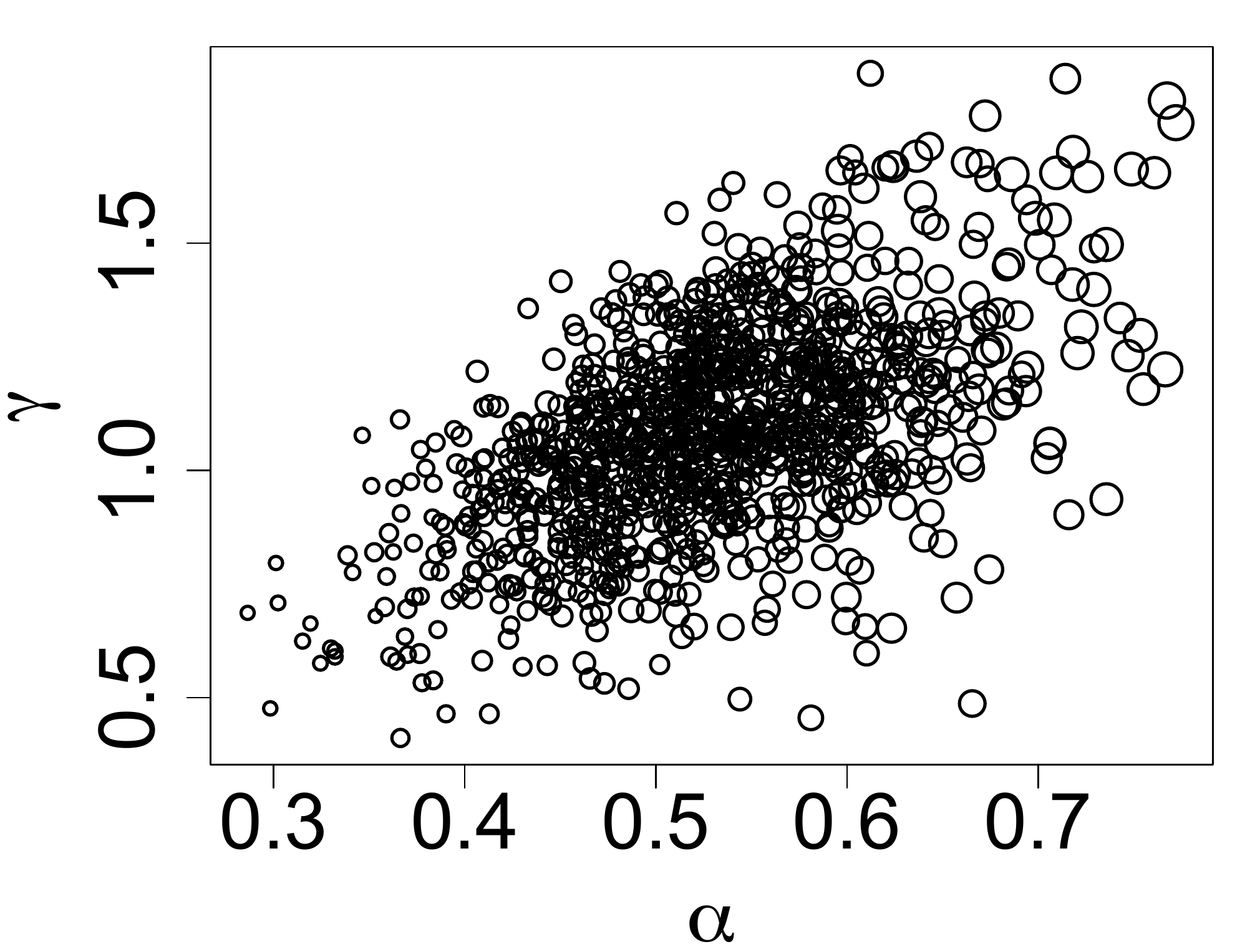} 
\caption{ABC Joint $(\alpha, \gamma)$}\label{subfig:joint_gamma_alpha_bate}
\end{subfigure} \\
\caption{
Pairwise joint ABC posteriors for astrophysical simulation data from \cite{Bate2012}.  Pairwise ABC posterior particles samples of (a) $(\lambda^{-1}, \alpha)$, (b) $(\lambda^{-1}, \gamma)$, and (c) $(\alpha, \gamma)$ for the astrophysical simulation data from \cite{Bate2012}.  The size of the plot symbol is scaled with the particle weight.
}
\label{fig:abc_bate_joints}
\end{figure}

\begin{figure}[htbp]
\centering
\includegraphics[width=.5\textwidth]{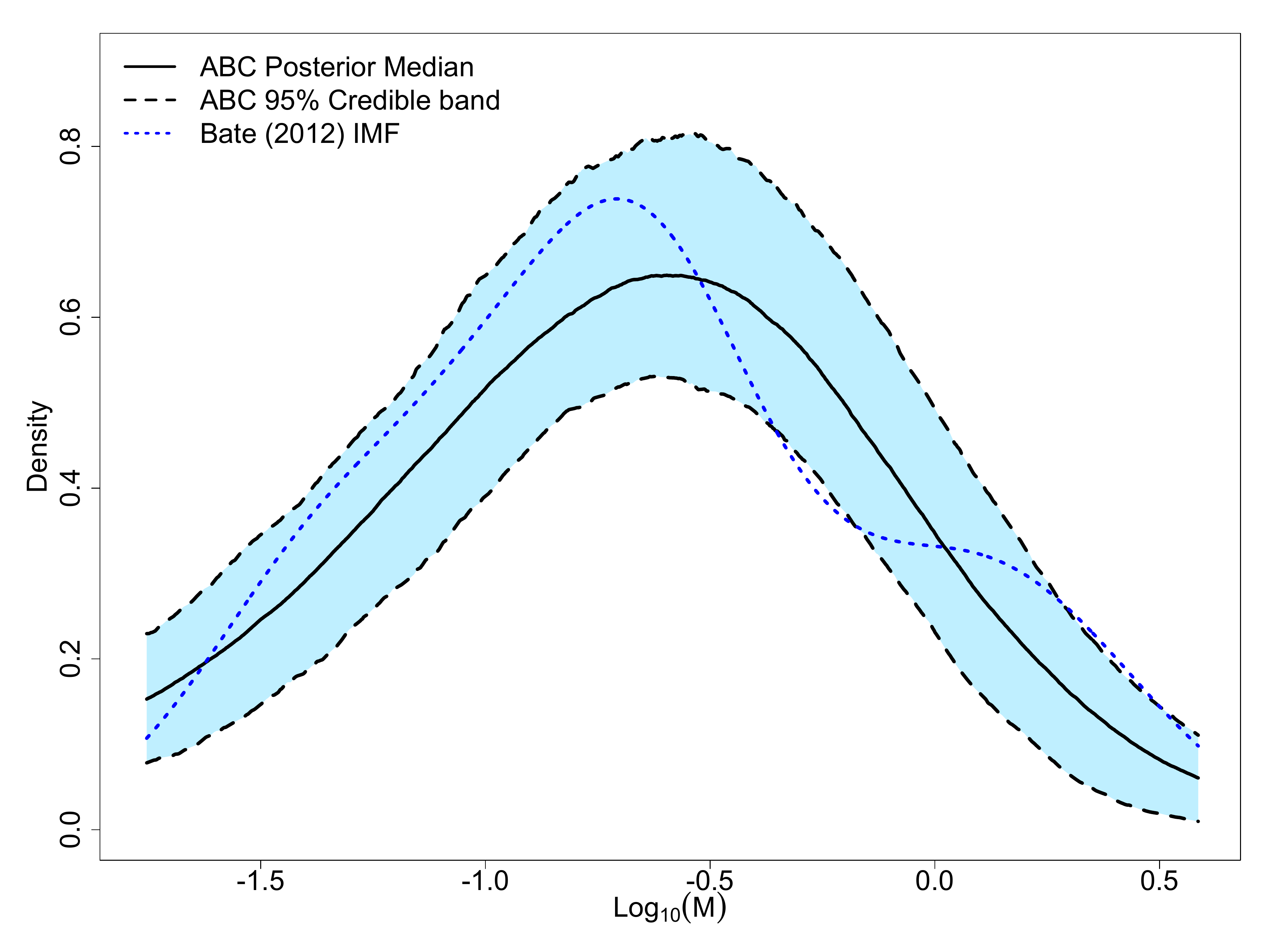}
 \caption{Posterior predictive IMF for astrophysical simulation data from \cite{Bate2012}. 
The median of the posterior predictive IMF (solid black) with a corresponding 95\% point-wise predictive band (dashed black) compared to the true IMF (blue dotted).  For the posterior predictive IMF, 1000 draws were made from the final ABC posteriors and then 1000 cluster samples were drawn from the proposed forward model.  
} \label{fig:abc_bate_pred}
\end{figure}

The 183 objects were used as the observations in the proposed ABC algorithm using 1000 particles, 5 sequential time steps, a $kN$ of $10^4$ (for adaptively initializing the algorithm), and the 25th percentile for shrinking the sequential tolerances based on the empirical distribution of the retained distances from the preceding time step.  The resulting ABC marginal posteriors are displayed in Figure~\ref{fig:abc_bate_posterior}, the pairwise ABC joint posteriors in Figure~\ref{fig:abc_bate_joints}, and the posterior predictive IMF in Figure~\ref{fig:abc_bate_pred}.
The ABC posterior means for $\lambda^{-1}$, $\alpha$, and $\gamma$ are 0.260, 0.537, and 1.091, respectively.  The ABC posterior mean of $\alpha$ is notably higher than the ABC posterior means of $\alpha$ for the \cite{kroupa2001} (0.293) and \cite{Chabrier:2003oq, Chabrier:2003om} (0.304) simulated data discussed in Section~\ref{sec:pa} (see Figure~\ref{fig:otherModels}).  The ABC posterior mean of $\gamma$ is also slightly higher than the 1.050 posterior mean of the \cite{Chabrier:2003oq, Chabrier:2003om} data.
Though the IMF has a slightly irregular shape with a small second mode around 1 $\Msun$ as noted previously, the proposed ABC method's  posterior predictive median and 
95\% predictive bands generally fit the IMF shape well.

\section{Discussion}
\label{discussionSec}

Accounting for the complex dependence structure in observable data, such as the initial masses of stars formed from a molecular cloud, 
is a challenging statistical modeling problem.  A possible, but unsatisfactory, resolution is to proceed as though the dependency 
is sufficiently weak that an independence assumption is acceptable. Such approximations can be reasonable at small sample sizes, but
are often revealed to be insufficient by modern data sets.
Instead, we draw on PA models, proposing a new forward model for star formation.
Though the new generative model was motivated by inference on stellar IMFs, the general concept is
generalizable to other applications. Simulation-based approaches to inference, including ABC, allow for inference
with such models.

The new generative model starts with the total mass of the system and stochastically builds individuals stars of particular mass at a sub-linear, linear, or super-linear rate.  
A goal of the proposed model and algorithm is to begin making a statistical connection between the observed stellar MF and the formation mechanism of the cluster, not that the proposed model shape is superior to the standard IMF models.  Rather, the proposed model is more general in the sense that it 
captures the dependencies among the masses of the stars by connecting the star masses to a possible cluster formation mechanism, and also can accommodate standard models proposed in the astronomical literature.
Additionally, by coupling the proposed model with ABC, observational limitations such as the aging and completeness of the observed cluster can easily be accounted for.
Code for running the proposed ABC-IMF algorithm is available at \url{https://github.com/JessiCisewskiKehe/ABC_for_StellarIMF}.

\new{In agreement with other studies that have implemented ABC algorithms (e.g. \citealt{WeyantEtAl2013,IshidaEtAl2015}), we found the selection of informative summary statistics to be a crucial, but challenging step in the algorithm development.  In the IMF setting, we had initially considered a number of different possible summary statistics, but it became clear that matching the shape of the IMF was important to constrain the parameters (along with the number of stars generated in the cluster).  
To assess the similarity between the observed and simulated IMFs, the $L_2$ distance was effective, but we imagine that other functional distances could also work well.  In future applications of ABC, practitioners may find it useful to consider functional summaries and distances if the setting allows for it.
To reach these conclusions, it required us to create a simplified setting where the true posterior was available; when possible, we suggest others consider this when trying to select useful summary statistics and distance functions.
}

While the proposed model is able to account for a particular dependency among the masses during cluster formation, there are several extensions that would be scientifically and statistically interesting.   First, the generative model could be extended to capture the spatial dependency among the observations.
Intuitively, such an approach could account for a mechanism that limits the formation of multiple very massive stars relatively near to each other.  
Understanding the spatial distribution of masses of stars during formation would help advance our understanding of stellar formation and evolution.  
Other effects that could be incorporated into the generative model include accounting for binary and other multiple star systems, the possible disturbances to the observed MF as stars die (beyond the censoring of the most massive stars), or spatial completeness functions (i.e. a completeness function that depends on not only the mass of the object, but also its location in the cluster).  

Hence, the proposed generative model, used in conjunction with ABC, provides a useful framework for 
dealing with complex physical processes that are otherwise difficult to work with in a 
statistically rigorous fashion. As increasing computational resources allow for greater model 
complexity in astronomy and other fields, the proposed and other ABC 
algorithms may open new opportunities for Bayesian inference in challenging problems. There appears to
be significant potential to extend this approach to even more complicated situations.

\appendix

\section{Generating power law tails} \label{sec:powerlaw}

As \remove{noted above} \new{mentioned in Section~\ref{sec:background}}, the PA model with linear evolution (the Yule-Simon Process) is 
known to generate power law tails \citep{newman2005}.
It is worth exploring the extent to which power law tail behavior is present 
in cases where $\gamma \neq 1$, as the power law model is a prevalent
assumption in this application\remove{.}\new{,  such as with the \cite{kroupa2001} and \cite{Chabrier:2003oq, Chabrier:2003om} models.}
For example, it would be of interest to determine
if tests of $H_0\!:\gamma = 1$ would have power to detect deviation from
power law tails\remove{.}\new{, which would be of interest to astronomers.}

\begin{figure}[htbp]
   \centering
\includegraphics[width = \textwidth]{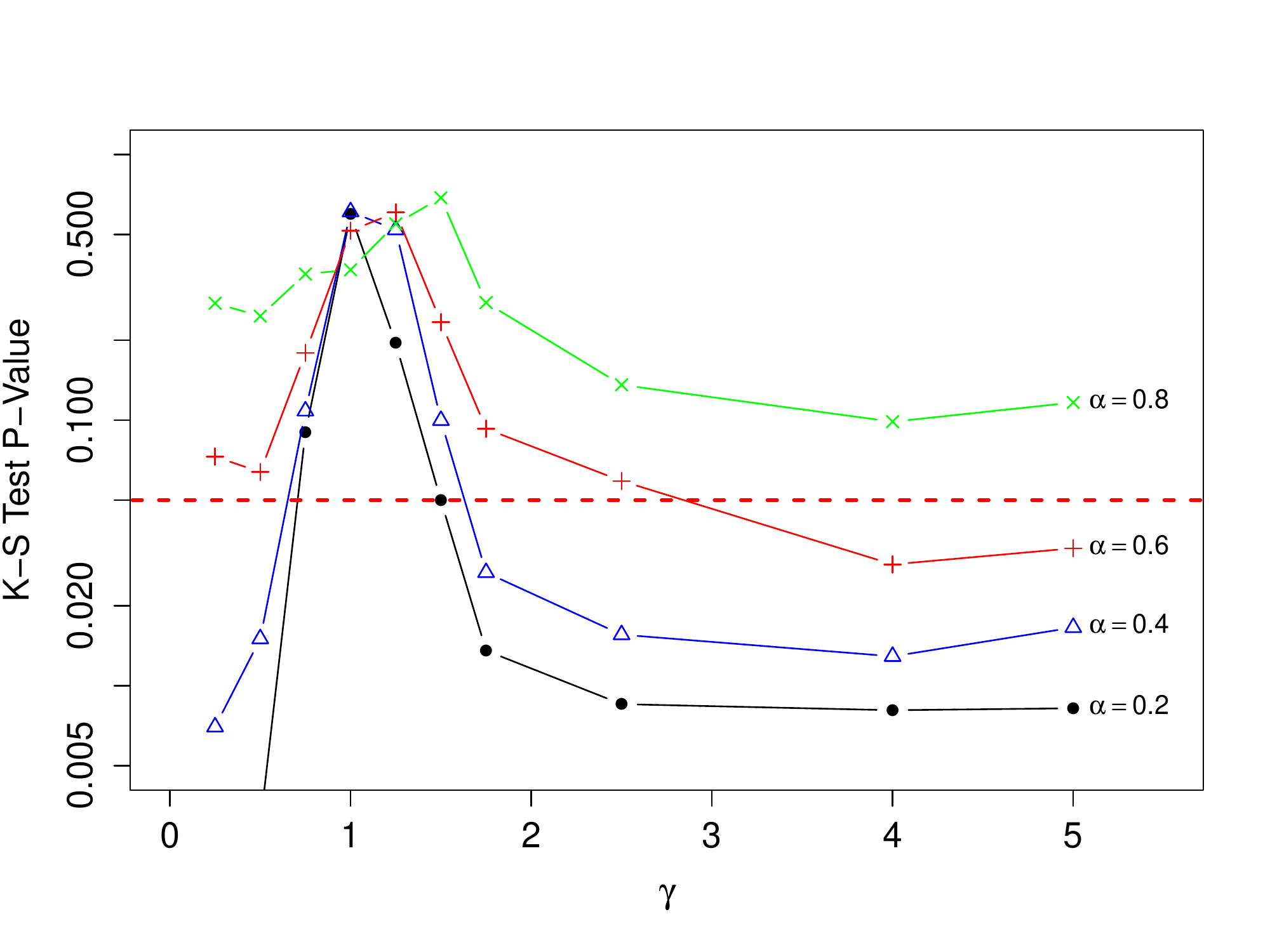}
\caption{Median p-values from Kolmogorov-Smirnov tests when comparing the tail distribution of masses simulated from
the proposed PA model with the best-fitting power law model. Fifty repetitions are done at each $(\alpha, \gamma)$
combination. The 0.05 cutoff is shown as a guide. Note that the vertical axis is on the log-scale.
The first two $\alpha = 0.2$ p-values drops below the range of the vertical axis to $9.392\times10^{-5}$ and $2.991\times10^{-3}$.
}
   \label{fig:KSpowerlaw1000}
\end{figure}

A small simulation study was conducted. Goodness-of-fit was assessed using the
standard Kolmogorov-Smirnov statistic, with the empirical distribution
of the masses of a collection of stars generated 
from our PA model compared to the best fitting power law model.
As we are only interested in fitting to the upper tail, this analysis 
is restricted to the region above $1 \Msun$. We fix $\lambda^{-1} = 0.25$
and $\Mtot = 1000$, and consider $\alpha \in \{0.2,0.4,0.6,0.8\}$, for
values of $\gamma$ ranging from 0.25 to 5. Fifty data sets are generated
for each $(\alpha, \gamma)$ combination. Results are shown in Figure \ref{fig:KSpowerlaw1000}.
In order to place the goodness-of-fit on a readily-interpretable scale, the p-value
is calculated for each K-S test, and the median across the 50 repetitions is displayed.
The results support the claim that the tail follows the power law when $\gamma = 1$, but that
the power law fit degrades quickly for $\gamma$ outside $(0.5, 1.5)$. The effect is particularly
strong for smaller values of $\alpha$. \new{In practice, one may decide to use a prior distribution on $\gamma$ that places more mass within
$(0.5, 1.5)$ if a power law model is expected, instead of the uniform prior distribution considered in this paper.} 


\section{Proposed ABC algorithm} \label{methodSec:abc}

The proposed ABC algorithm is displayed in Algorithm~\eqref{alg1}, where $N$ is the desired particle sample size to approximate the posterior distribution, and is motivated by the adaptive and sequential ABC algorithm of \cite{beaumont2009}.  
The forward model, $F$, in Algorithm~\eqref{alg1} is where the IMF masses are drawn and observational limitations and uncertainties, stellar evolution, and other astrophysical elements can be incorporated as outlined in Section~\ref{PAmodelSection}.  The other details of the proposed algorithm are discussed next.

Algorithm~\eqref{alg1} is initialized using the basic ABC rejection algorithm at time step $t = 1$ using a distance function $\rho(\msim, \mobs)$ to measure the distance between the simulated and observed datasets, $\msim$ and $\mobs$, respectively.  The first tolerance, $\epsilon_1$, is adaptively selected by drawing $kN$ particles for some $k >0$.  Then the $N$ particles that have the smallest distance are retained, and $\epsilon_1$ is defined as the largest of those $N$ distances retained.
For subsequent time steps ($t > 1$), rather than proposing a draw, $\theta^*$, from the prior, $\pi(\theta)$, the proposed $\theta^*$ is selected from the previous time step's ($t-1$) ABC posterior samples.  The selected $\theta^*$ is then moved according to some kernel, $K(\theta^*, \cdot)$, to ameliorate degeneracy as the sampler evolves.  In order to ensure the true posterior (which requires sampling from the prior) is targeted, the retained draws are weighted according to the appropriate importance weights, $W_t$ -- this incorporates the proposal distribution's kernel.

\begin{algorithm}[ht]
 \caption{Stellar IMF ABC algorithm with sequential sampling} \label{alg1}
 \KwData{Observed stellar masses, ($\mobs$)}
 \KwResult{ABC posterior sample of $\theta$}

 {\em At iteration $t = 1$:}\\
  \For{$j = 1, \ldots, kN$}{
 Propose $\theta^{*(j)}$ by drawing from $\pi(\theta)$ \\
 Generate cluster stellar masses $\msim$ and apply other effects from $F(m \mid \theta^{*(j)})$ \\
Calculate distance $\rho_t^{(j)} \gets \rho(\msim, \mobs)$
 }
 
$\theta_t^{(j)} \gets \theta^{*(l)}, l = $ indices of $N$ smallest $\rho_t^{(q)}, q = 1, \ldots, kN$\\
$\epsilon_{t+1} \gets$ desired quantile of $\rho_t^{(l)}$ with $l$ defined as above\\
$W_t^{(j)} \gets 1/N,  j = 1, \ldots, N$ \\

{\em At iterations $t = 2, \ldots, T$:}\\
 \For{$j = 1, \ldots, N$}{
 \While{$\rho^{*(j)} > \epsilon_t$}{
 Select $\theta^{(j)}$ by drawing from the $\theta^{(i)}_{t-1}$ with probabilities $W^{(i)}_{t-1}, i = 1, \ldots, N$ \\
 Generate $\theta^{*(j)}$ from transition kernel $K(\theta^{(j)}, \cdot)$ \\
 Generate cluster stellar masses $\msim$ and apply other effects from $F(m \mid \theta^{*(j)})$ \\
 Calculate distance $\rho^{*(j)} \gets \rho(\msim, \mobs)$
 }
$\theta_t^{(j)} \gets \theta^{*(j)}$, \quad $\rho_t^{(j)} \gets \rho^{*(j)}$\\
$W_t^{(j)} \gets \frac{\pi(\theta_t^{(j)})}{\sum_{i = 1}^N  W_{t-1}^{(i)} K(\theta_{t-1}^{(i)}, \theta_t^{(j)})} $
}
$W_t^{(j)} \gets \frac{W_t^{(j)}}{\sum_{l = 1}^N W_t^{(l)}}$, \quad $\epsilon_{t+1} \gets$ desired quantile of $\rho_t^{(j)}, j = 1, \ldots, N$\\
 \bigskip
\end{algorithm}

A key step in the implementation of an ABC algorithm is to quantify the distance between the simulated and observed stellar masses.  We define a bivariate summary statistic and distance function that captures the shape of the present-day MF and the random number of stars observed, displayed in Equations~\eqref{eq:2da} and \eqref{eq:2db}, respectively.  For the shape of the present-day MF, we use a kernel density estimate of the $\log_{10}$ masses (due to the heavy-tailed distribution of the initial masses), and an $L_2$ distance between the simulated and observed $\log_{10}$ MF estimates.  
The number of stars observed is the other summary statistic, with the distance being the absolute value of the difference in the ratio of the counts from 1. More specifically, the bivariate summary statistic is defined as
\begin{align}
\rho_1(\msim, \mobs) &= \left [\displaystyle \int \left \{\hat f_{\log \msim}(x) - \hat f_{\log \mobs}(x) \right \}^2 dx \right]^{1/2} \label{eq:2da} \\ 
\rho_2(\msim, \mobs) &= \max\left\{\left|1 - \nsim/ \nobs\right |, \left|1 - \nobs/ \nsim\right |  \right\} \text{,}   \label{eq:2db}
\end{align}
where the $\hat f_z$ are kernel density estimates of $z$, and $\nsim$ and $\nobs$ are the number of stars comprising the simulated and observed MF, respectively.  These summary statistics were selected based on performance of a simulation study using the high-mass section of the broken power-law model because the true posterior is known in this setting.  Results and additional discussion of the simulation study \remove{can be found in} \new{is presented below in} Appendix~\ref{app:summary}.

With the bivariate summary statistic, we use a bivariate tolerance sequence, $(\epsilon_{1t}, \epsilon_{2t})$, for $t = 1, \ldots, T$ is such that $\epsilon_{i1} \geq \epsilon_{i2} \geq \cdots \geq \epsilon_{iT}$ for $i = 1, 2$.  At time step $t$, the tolerances are determined based on the  empirical distribution of the retained distances from time step $t-1$ (e.g. the $25th$ percentile).  As noted previously, the tolerance sequence is initialized adaptively by selecting $kN$ proposals from the prior distributions, then the $N$ proposals that result in the $N$ smallest distances were selected.\footnote{The $kN$ sampled distances were scaled, squared, and then added together; the $N$ smallest of these combined distances were retained.}
The distance function and tolerance sequence displayed in Algorithm~\eqref{alg1} are defined as $\rho_t(\msim, \mobs) = \{\rho_{1t}(\msim, \mobs), \rho_{2t}(\msim, \mobs)\}$, and $\epsilon_t = \{\epsilon_{1t}, \epsilon_{2t}\}$ (which can also be expanded to include the additional summary statistic noted below).

In practice, $\Mtot$ is an unknown quantity of interest.  A prior can be assigned to $\Mtot$ and an additional summary statistic and tolerance sequence can be used.  The summary statistic selected in this case is
\begin{equation}
\rho_3(\msim, \mobs) = \left| \sum_{i = 1}^{\nsim} m_{\text{sim}, i} - \sum_{j = 1}^{\nobs} m_{\text{obs}, j} \right|, \label{eq:2dc}
\end{equation}
where $m_{\text{sim}, i}$ and $m_{\text{obs}, j}$ are the masses of the individual simulated and observed stars, respectively.

\subsection{ABC summary statistic selection} \label{app:summary}
In order to select effective summary statistics for the proposed model, we first employ the ABC methodology in a simplified study that focuses on the posterior of the power law parameter $\alpha$ from Equation~\eqref{eq:imf}.  We generate a cluster of $n = 10^3$ stars from an IMF with slope $\alpha = 2.35$  \citep{salpeter55}, $\Mmin = 2$, and $\Mmax = 60$,  and a uniform prior distribution for $\alpha \in (0, 6)$.  
This  model was used in order to check the method against the true posterior of $\alpha$ after the observational and aging effects have been incorporated into the forward model.  We use the bivariate summary statistic and distance function of Equation~\eqref{eq:2da} and \eqref{eq:2db}.  Defining the two-dimensional tolerance sequence as $(\epsilon_{1t}, \epsilon_{2t})$ where the subscript $t$ indicates the algorithm time step, and  $\epsilon_{11}$ and $\epsilon_{21}$ were selected using an adaptive start as discussed \remove{in Section}\new{above} using an initial number of draws of $10N$ with $N = 10^3$.  The algorithm ran for $T = 5$ time steps.
At steps $t = 2,\ldots,T$, $\epsilon_{1t}$ and $\epsilon_{2t}$ were set equal to the $25th$ percentile of the distances retained at the previous step from their corresponding distance functions.

\begin{figure}[htbp]
\centering
\includegraphics[width=.5\textwidth]{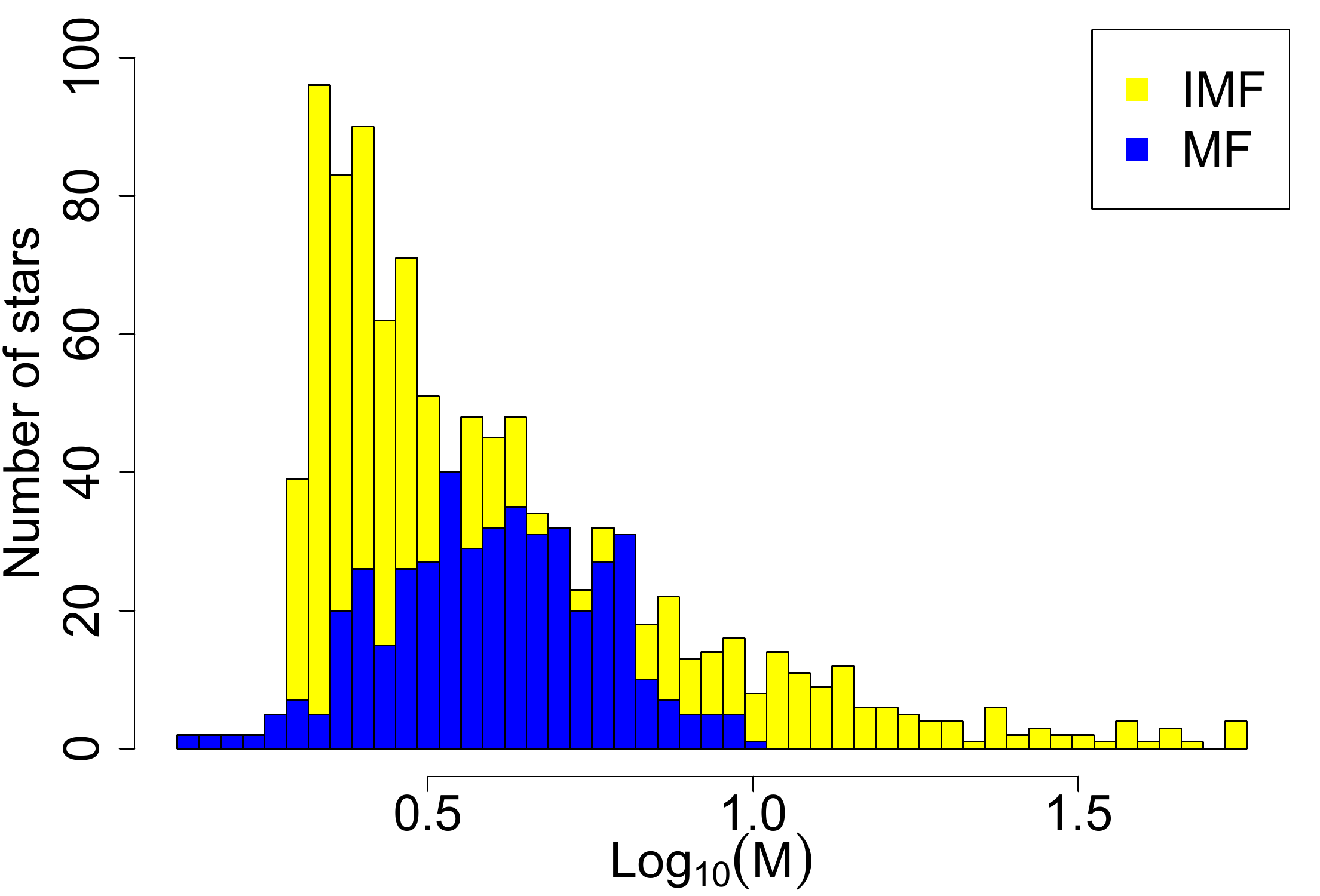}       
\caption{Simulated IMF (yellow) and MF (blue) using power law model.  The IMF was simulated with $n = 10^3$ stars using a power law slope of $\alpha = 2.35$.  The cluster was aged 30 Myrs, simulated with log-normal measurement error with $\sigma = 0.25$, and had a linear-ramp completeness function applied between 2 and 4 $\Msun$.
} \label{fig:abc_simple_data}
\end{figure}

The pseudo-data were aged 30 Myr, log-normal measurement error with $\sigma = 0.25$, and observation completeness defined by the linear-ramp function in \eqref{eq:ramp} with $\Cmin = 2$ $\Msun$ and $\Cmax = 4$ $\Msun$.  The simulated IMF and resulting MF (after the noted observational effects were applied) are displayed in Figure~\ref{fig:abc_simple_data}.  The IMF is the object of interest, while the MF contain the actual observations that can be used for analysis.

\begin{figure}[htbp]
\begin{subfigure}{0.32\textwidth}
\centering
\includegraphics[width=\textwidth]{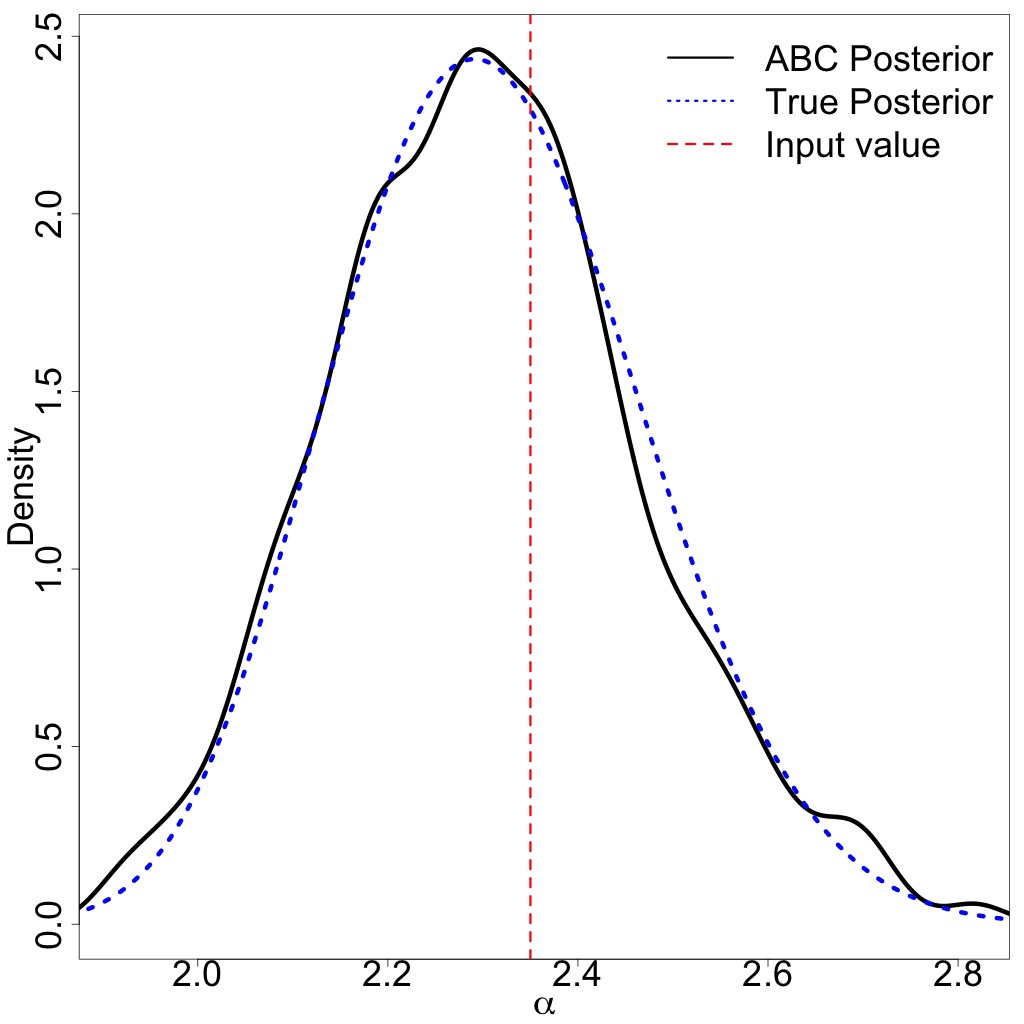}
\caption{Psosterior}\label{subfig:basic_alpha}
\end{subfigure}
\begin{subfigure}{0.32\textwidth}
\centering
\includegraphics[width=\textwidth]{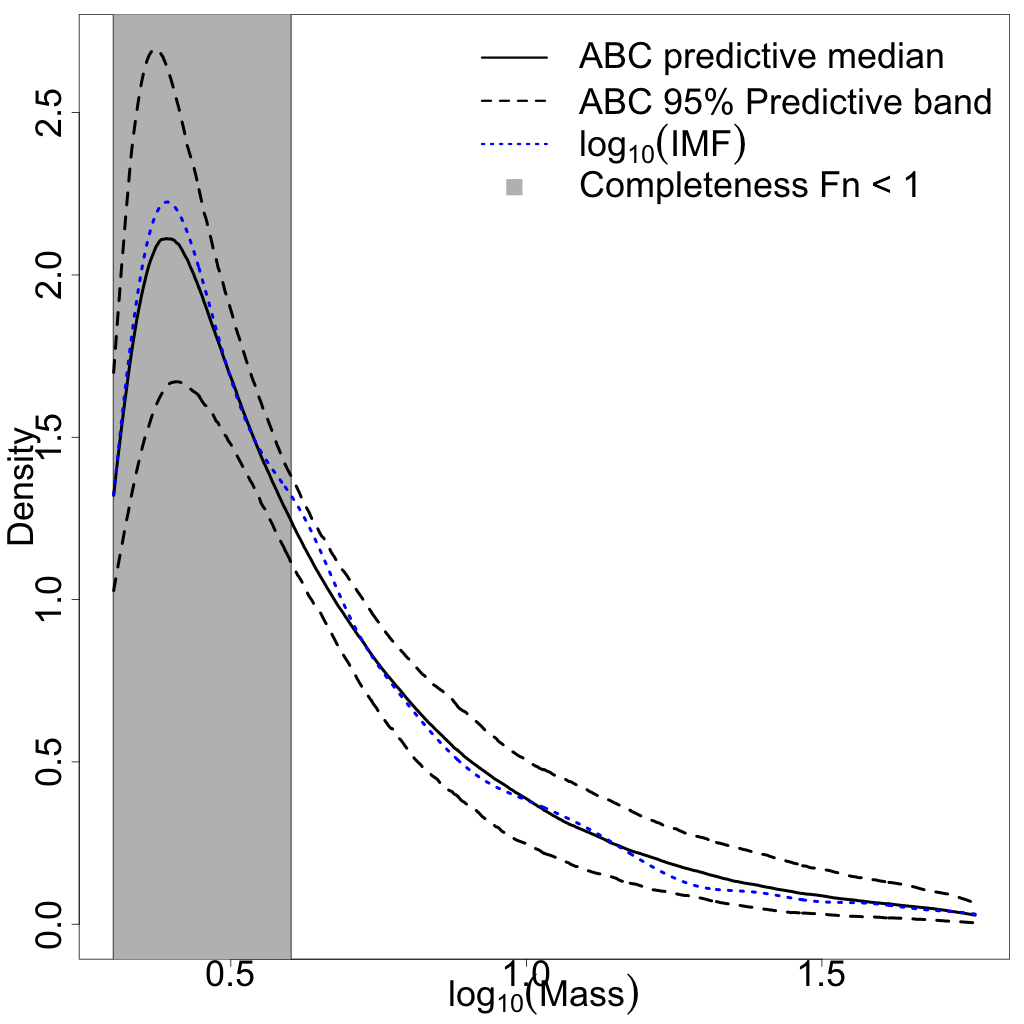}
\caption{Predictive IMF}\label{subfig:basic_imf}
\end{subfigure}
\begin{subfigure}{0.32\textwidth}
\centering
\includegraphics[width=\textwidth]{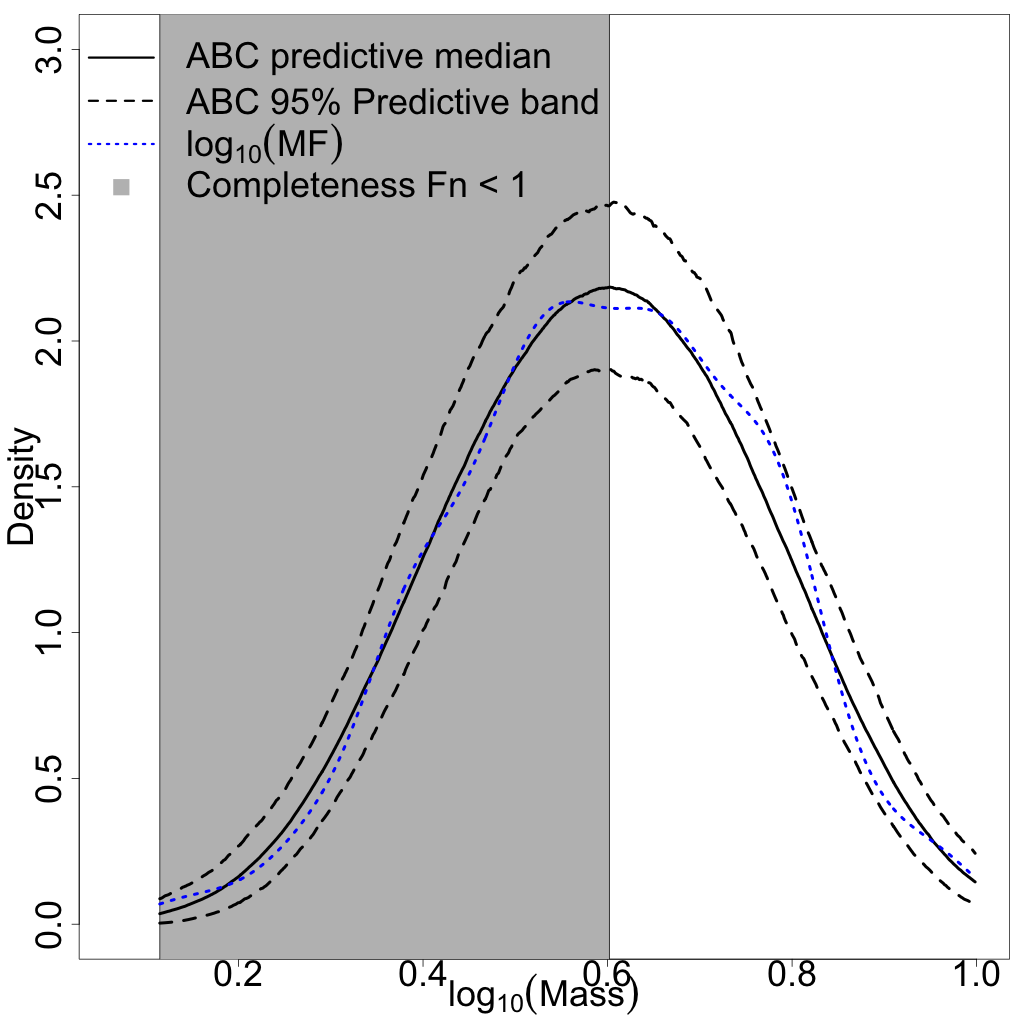}
\caption{Predictive MF}\label{subfig:basic_mf}
\end{subfigure}        
   \caption{Validation of summary statistics with power law model.  (a) The ABC posterior for $\alpha$ (solid black) compared to the true posterior (dotted blue) of Equation~\eqref{eq:simple_posterior} using an input value of 2.35 (dashed vertical red).  (b) The median of the posterior predictive IMF (solid black) with a corresponding 95\% point-wise predictive band (dashed black) compared to the true IMF (blue dotted) which was the simulated dataset before aging, completeness, or uncertainty were applied, and the gray shaded region indicates where the completeness function was less than 1.  (c) The median of the posterior predictive MF (solid black) with a corresponding 95\% point-wise predictive band (dashed black) compared to the observed MF (dotted blue) which was the simulated dataset after aging, completeness, and uncertainty were applied.  For the posterior predictive IMF, 1000 independent draws were made from the ABC posterior of (a) and then 1000 cluster samples were drawn from the power law simulation model.  For the posterior predictive MF, the 1000 cluster samples used for (b) were then put through the forward model to apply the aging, completeness, and measurement error effects.
} \label{fig:abc_simple}
\end{figure}

The ABC posterior resulting from the ABC algorithm along with the true posterior for $\alpha$ are displayed in Figure~\ref{subfig:basic_alpha}.  The ABC posterior matches the true posterior, defined as 
\begin{align}
&\pi_F(\alpha \mid  \mobs, M_{\min}, M_{\max}, \nobs, n_{tot}, T_{age}) \propto  \label{eq:simple_posterior}\\ 
&\left \{\Proba(M>T_{age})+\left(\frac{1-\alpha}{M_{\max}^{1-\alpha}-M_{\min}^{1-\alpha}}\right)\int_2^4 M^{-\alpha}\left(1-\frac{M-2}{2}\right )dM\right \}^{n_{tot}-\nobs}  \nonumber \\ \nonumber
& \times \prod_{i=1}^{\nobs}\left\{ \int_2^{T_{age}}(2\pi\sigma^2)^{-\frac{1}{2}}m_i^{-1}e^{-\frac{1}{2\sigma^2}(\log(m_i)-\log(M))^2} \left(\frac{1-\alpha}{M_{\max}^{1-\alpha}-M_{\min}^{1-\alpha}}\right)M^{-\alpha} \right.\\ \nonumber
& \times  \left.\left(I\{M>4\}+\left(\frac{M-2}{2}\right)I\{2\leq M\leq 4\}\right) dM \right\} 
\end{align}
where $T_{age} = \tau^{-1/3} \times 10^{4/3}$ is the upper-tail mass cutoff due to aging.  The close match between the true and ABC posteriors suggests that the selected summary statistics are useful for carrying out the ABC analysis.  Figures~\ref{subfig:basic_imf} and \ref{subfig:basic_mf} display the ABC posterior predictive IMF and MF.  Even in regions where stars are missing due to the observational limitations, the ABC predictive median is still able to recover the shape of the original IMF (though with wider credible bands).

%
%
%
%

\section*{Acknowledgements}
\new{The authors thank the Yale Center for Research Computing for the resources we used for producing this paper.
The authors also thank two anonymous reviewers, along with David W. Hogg, and Ewan Cameron for their comments and feedback on this work.  }
Jessi Cisewski-Kehe and Grant Weller were partially supported by the National Science Foundation under Grant DMS-1043903. 
Chad Schafer was supported by NSF Grant DMS-1106956.   Any opinions, findings, and conclusions or recommendations expressed in this material are those of the authors and do not necessarily reflect the views of the National Science Foundation.


\bibliographystyle{imsart-nameyear}
\bibliography{imfABC.bib}

\end{document}